\documentclass[iop,twocolappendix,numberedappendix]{emulateapj}
\usepackage{epsfig}

\def\gtsima{$\; \buildrel > \over \sim \;$}
\def\ltsima{$\; \buildrel < \over \sim \;$}
\def\gtrsim{\lower.5ex\hbox{\gtsima}}
\def\lesssim{\lower.5ex\hbox{\ltsima}}

\shorttitle{RLO systems powered by BHs in young SCs}
\shortauthors{Mapelli \&{} Zampieri}

\begin{document}

\title{Roche-lobe overflow systems powered by black holes in young star clusters: the importance of dynamical exchanges}
\author{Michela Mapelli \&{} Luca Zampieri\altaffilmark{1}} %, E. Ripamonti$^{2}$, A. Bressan$^{3}$

%$^1$
\altaffiltext{1}{INAF-Osservatorio Astronomico di Padova, Vicolo dell'Osservatorio 5, I--35122, Padova, Italy; {\tt michela.mapelli@oapd.inaf.it}
%$^2$Universit\`a Milano Bicocca, Dipartimento di Fisica G. Occhialini, Piazza della Scienza 3, I--20126, Milano, Italy\\
%$^3$Scuola Internazionale Superiore di Studi Avanzati (SISSA), Via Bonomea 265, I--34136, Trieste, Italy
}
%\maketitle \vspace {7cm }

  \begin{abstract}
We have run 600 $N-$body simulations of intermediate-mass ($\sim{}3500$ M$_\odot{}$) young star clusters (SCs) with three different metallicities ($Z=0.01$, 0.1 and 1 Z$_\odot{}$). The simulations include the dependence of stellar properties and stellar winds on metallicity. Massive stellar black holes (MSBHs) with mass $>25$ M$_\odot{}$ are allowed to form through direct collapse of very massive metal-poor stars ($Z<0.3$ Z$_\odot$). We focus on the demographics of black hole (BH) binaries that undergo mass transfer via Roche lobe overflow (RLO). We find that 44 per cent of all binaries that undergo an RLO phase (RLO binaries) formed through  dynamical exchange. RLO binaries that formed via exchange (RLO-EBs) are powered by more massive BHs than  RLO primordial binaries  (RLO-PBs). Furthermore, the RLO-EBs tend to start the RLO phase later than the RLO-PBs. In metal-poor SCs ($0.01-0.1$ Z$_\odot{}$), $>20$ per cent of all RLO binaries are powered by MSBHs. The vast majority of RLO binaries powered by MSBHs are RLO-EBs. We have produced optical color-magnitude diagrams of the simulated RLO binaries, accounting for the emission of both the donor star and the irradiated accretion disk. We find that RLO-PBs are generally associated with bluer counterparts than RLO-EBs. We compare the simulated counterparts with the observed counterparts of nine ultraluminous X-ray sources. We discuss the possibility that IC~342 X-1, Ho~IX X-1, NGC~1313 X-2 and NGC~5204 X-1 are powered by a MSBH.
\end{abstract}

%
%________________________________________________________________

\section{Introduction}\label{sec:intro}

Most  stars ($\sim{}70-90$ per cent) are believed to form in star clusters (SCs, e.g. \citealt{lada03}; \citealt{porras03};  \citealt{parker07}; \citealt{portegies10}; \citealt{silva10}; \citealt{gvaramadze12}; \citealt{kharchenko12}; \citealt{kharchenko13}; see also \citealt{bressert10} and \citealt{gieles12} for a possible issue). 
%, since star formation (SF) takes place in the dense cores of giant molecular clouds (e.g. \citealt{bate09}, and references therein). 
The percentage of SCs that remain bound after the evaporation of gas (i.e. after the first few Myrs) is more uncertain: it may be as low as $\sim{}5$ per cent and as high as $\sim{}30$ per cent, and it might also depend on the environment (e.g. \citealt{bastian08}; \citealt{silva10}; \citealt{goddard10}; \citealt{gieles11}). The fraction of SCs that survive for $>100$ Myr is likely  lower ($\sim{}5$ per cent, \citealt{lada03}). %Claims che le stelle + massive si formano in scs.

%This  means that 
 Thus, most progenitors of stellar black holes (BHs) form in SCs, and a 
%conspicuous (although uncertain) 
number of BHs spend the first part of their life in a young SC, before being possibly ejected into the field. This is crucial to understand the formation and evolution of X-ray binaries.

In fact, dense young SCs are collisional stellar systems: their two-body relaxation timescale is shorter than (or comparable to) their lifetime (e.g. \citealt{portegies10}). For example, the half-mass relaxation timescale for a SC with total mass $M_{\rm TOT}\sim{}10^4\,{}{\rm M}_{\odot{}}$ and half-mass radius $r_{\rm hm}\sim{}1$ pc (such as NGC~3603 and Trumpler~14) is $t_{\rm rlx}\sim{}20$ Myr (e.g. \citealt{portegies02}). This implies that close encounters between stars and binaries (three-body encounters), and even triples or multiple systems (e.g.~\citealt{leigh13}) play an important role in the overall dynamical evolution of a SC, as well as in the fate of single objects (e.g. \citealt{hills75}). During a three-body encounter, the binary and the single star exchange energy (e.g. \citealt{heggie75}). This alters the orbital properties of the binary, and may induce recoil on the center of mass of the involved bodies (e.g. \citealt{aarseth72}; \citealt{hills80}; \citealt{heggie93}; \citealt{sigurdsson93}; \citealt{sigurdsson95}; \citealt{davies95}; \citealt{colpi03}; \citealt{mapelli05}). Dynamical exchanges can also occur, i.e. three-body encounters during which the single star replaces one of the former members of the binary. 
%: during a three-body encounter, the single star replaces one of the former members of the binary. 
The probability of an encounter to end up with a dynamical exchange is higher if the mass of the single star is equal or higher than the mass of one of the binary members (\citealt{hills89}; \citealt{hills92}). %{\bf Multiple encounters can also occur, involving exotic systems such as hyerarchical triples

These dynamical effects are crucial %for the formation of stellar exotica (Davies), such as blue straggler stars (). They are also important 
for the formation of BH binaries (i.e. binaries hosting at least one BH): a BH that was born single in the field will likely remain single forever, whereas a single BH  in a collisional SC may acquire a companion through dynamical exchanges. Since the probability of exchange depends on the mass, this process will favor the formation of BH binaries hosting the most massive BHs in a SC (e.g. \citealt{hills76}; \citealt{hills91}). \citealt{mapelli13} (2013, hereafter M13) studied the dynamical evolution of BHs in young SCs with various metallicity, and found that $20-25$ per cent of simulated BHs form from single stars and become members of binaries through dynamical exchange in the first 100 Myr of the SC life. This fraction rises to $\sim{}75$ per cent if only the most massive BHs ($>25$ M$_\odot$) are considered. Also, dynamical encounters induce a mass-transfer phase and the birth of an X-ray binary.
%Only a fraction of   ---> important for X-ray binaries because all progenitors of BHs and NSs form in SCs.
%-LADA
%-portegies
%-kharchenko 2012, 2013
%-gieles, moeckel & clarke
%-Gvaramadze, V. V.; Weidner, C.; Kroupa, P.; Pflamm-Altenburg, J. 2012
%	Monthly Notices of the Royal Astronomical Society, Volume 424, Issue 4, pp. 3037-3049.

From an observational perspective, bright X-ray binaries and ultraluminous X-ray sources (ULXs, i.e. point-like non-nuclear X-ray sources with luminosity, assumed isotropic, $L_X>10^{39}$ erg s$^{-1}$) are often associated with OB associations and with young  SCs (e.g. \citealt{goad02};  \citealt{zezas02}; \citealt{liu04}; \citealt{soria05}; \citealt{ramsey06}; \citealt{terashima06}; \citealt{abolmasov07}; \citealt{berghea09}; \citealt{swartz09}; \citealt{tao11}; \citealt{grise11};  \citealt{grise12}; \citealt{bodaghee12}; \citealt{coleiro13}). This may indicate that the dynamics of the parent SC enhances the formation of bright X-ray binaries. In addition, a significant fraction of these bright X-ray sources are found to be slightly displaced (by few tens to hundreds of parsecs) from the closest young SC (\citealt{zezas02}; \citealt{kaaret04}; \citealt{berghea09}; \citealt{rangelov12}; \citealt{poutanen13}; \citealt{berghea13}). This has been interpreted as a consequence of natal kicks (e.g. \citealt{sepinsky05}; \citealt{zuo10}) and of dynamical recoil (e.g. \citealt{kaaret04}; \citealt{mapelli11b}).

Population synthesis simulations of field binaries provide valuable information about the demographics of X-ray binaries (e.g. \citealt{portegies97}; \citealt{hurley02}; \citealt{podsiadlowski02}; \citealt{podsiadlowski03}; \citealt{belczynski04a}; \citealt{belczynski04b}; \citealt{rappaport05}; \citealt{dray06}; \citealt{madhusudhan06}, 2008; \citealt{belczynski08}; \citealt{linden10}), but do not take into account the effects of dynamics. The few studies of BH demographics in young SCs that include both stellar evolution and dynamics (e.g. \citealt{blecha06}; M13; \citealt{goswami13}) highlight that the effects induced by dynamics (especially dynamical exchanges) cannot be neglected, even in the first 10 Myr of the young SC life. In particular, M13 is the first study of the dynamical evolution of BHs in young SCs that includes self-consistent recipes for the formation of massive stellar BHs (MSBHs, with mass $m_{\rm BH}>25$ M$_\odot$) in metal-poor environments. In this paper, we adopt the same recipes as in M13, and we focus on the demographics of BH binaries that undergo Roche lobe overflow (RLO). We give particular attention to the formation pathways of RLO systems and to the properties of the donor star.

%main difference with field is dynamics....cfr pop synthesis with 

%exchanges!!!

%in paper I.. here we focus on RLO and double the sample

The paper is structured as follows. In Section 2,  we briefly describe the method adopted for the simulations.  In Section 3, we present the results, focusing on the formation pathways of BHs (Section 3.1), on the importance of dynamical exchanges  for RLO binaries (Section 3.2), on the role of  MSBHs (Section 3.3), on the properties of the donor stars (Section 3.4), and on the distribution of the simulated binaries in the color-magnitude diagram (CMD, Section 3.5).  In Section 4, we discuss the main results and compare the simulated systems with the observed counterparts of some ULXs. In Section 5, we summarize the most relevant results and discuss future challenges for $N-$body simulations of SCs.

%In the following, we briefly review the previous work on this topic, to summarize the state-of-art and to highlight the differences with the current paper. }

\section{Method and simulations}\label{sec:sec2}
The simulations have been done using the {\sc starlab} public software environment (\citealt{portegies01}), which %follows the dynamical evolution of a SC, resolving binaries and three-body encounters. {\sc starlab} 
includes the {\sc kira} direct-summation $N$-body integrator and the {\sc seba} code for stellar and binary evolution (\citealt{portegies96}; \citealt{portegies01}; \citealt{nelemans01}). The recipes for binary evolution adopted in this paper are the same as described in \citet{portegies01}, while the recipes for stellar evolution  are the same as described in M13 (see also \citealt{mapellibressan13}). In particular, {\sc seba} was  modified in M13, to include various effects of metallicity, as follows.
%\footnote{The original version of {\sc seba} accounts only for a solar-metallicity environment.}, as follows.

In M13,  the metallicity dependence of stellar radius, temperature and luminosity was added to {\sc seba}, using the polynomial fitting formulas by \citet{hurley00}.  The recipes for mass loss by stellar winds were updated, by using the metal-dependent fitting formulas  for main sequence (MS) stars provided by \citealt{vink01} (2001; see also \citealt{belczynski10}). Furthermore, we added an approximate treatment of stellar winds for luminous blue variable (LBV) stars and for Wolf-Rayet (WR) stars.  A post-MS star becomes a LBV star if $L/{\rm L}_\odot{}>6 \times 10^5$ and $10^{-5}\,{}(R/{\rm R}_\odot{})\,{}(L/{\rm L}_\odot{})^{0.5} >1.0$, where $L$ and $R$ are the luminosity and the radius of the star, respectively (\citealt{humphreys94}). A LBV star loses mass by stellar winds at a rate $\dot{M} = f_{\rm LBV} \times 10^{-4}$ M$_\odot{}$ yr$^{-1}$, where $f_{\rm LBV}=1.5$ (\citealt{belczynski10}). Naked helium-giant stars with zero-age MS (ZAMS) mass $m_{\rm ZAMS}>25$ M$_\odot{}$ are considered WR stars (e.g. \citealt{vanderhucht91}), and lose mass by stellar winds at a rate\footnote{In this definition and throughout the text, we adopt Z$_{\odot{}}=0.019$.} $\dot{M} = 10^{-13} (L/{\rm L}_\odot{})^{1.5}\,{}({Z/{\rm Z}_\odot})^{\beta{}}$ M$_\odot{}$ yr$^{-1}$, where $\beta{}=0.86$ (\citealt{hamann98}; \citealt{vink05}; \citealt{belczynski10}).

 A star with ZAMS mass $8\le{}m_{\rm ZAMS}/{\rm M}_\odot{}<25$ undergoes supernova (SN) explosion by the end of its life and becomes a neutron star\footnote{We assume that all stars with $8\le{}m_{\rm ZAMS}/{\rm M}_\odot{}<25$ ($m_{\rm ZAMS}\ge{}25\,{}{\rm M}_\odot{}$) become NSs (BHs), regardless of metallicity. The actual threshold is expected to depend on metallicity. On the other hand, our assumption is quite robust up to $Z\sim{}{\rm Z}_\odot{}$ (see e.g. figure~1 of \citealt{heger03}).} (NS). The NS receives a natal kick randomly selected from the distribution $P(u)=\frac{4}{\pi{}}\,{}\frac{1}{(1+u^2)^2}$, where $u=v/\sigma{}$ ($v$ is the NS velocity modulus and $\sigma{}=600$ km s$^{-1}$, \citealt{hartman97}).

Stars with  ZAMS mass $m_{\rm ZAMS}\ge{}25\,{}{\rm M}_\odot{}$ and final mass (i.e. the mass bound to a star immediately before the collapse) $m_{\rm fin}<40$ M$_\odot$ undergo SN explosion and become a BH, because of fallback of material from the SN ejecta. The natal kick for a BH born by SN explosion is drawn from the same distribution as that of NSs, but  scaled to $m_{\rm NS}/m_{\rm BH}$ (where $m_{\rm NS}$ is the typical NS mass, here approximately taken to be equal to 1.34 M$_\odot{}$).

Stars with final mass  $m_{\rm fin}\ge{}40$ M$_\odot$ collapse to a BH directly, without SN explosion (\citealt{fryer99}). The mass of a BH born from direct collapse is very similar to $m_{\rm fin}$, since there are no ejecta. In particular, the mass spectrum of BHs born from direct collapse is the same as described in Fig.~1 of M13. Since our recipes of mass loss by stellar winds depend on metallicity (\citealt{vink01} 2001; \citealt{vink05}), metal-poor stars have higher  values of $m_{\rm fin}$ than metal-rich stars with the same $m_{\rm ZAMS}$, and collapse to more massive BHs. BHs with mass up to $\sim{}40$ (80) M$_\odot{}$ can form at $Z=0.1$ (0.01) Z$_\odot$ by direct collapse of single stars. Furthermore, BHs born from direct collapse do not receive any natal kicks (\citealt{fryer12}).

In our simulations, BHs with mass $>25$ M$_\odot$ are allowed to form even through merger of a BH or a NS with another star. We assume no mass loss during the merger (which is rather optimistic, e.g. \citealt{gaburov10}), since we do not have recipes for hydrodynamical treatment of the star-star mergers. BHs with mass $\ge{}25$ M$_\odot{}$ (which form either by direct collapse or by merger) are named massive stellar BHs (MSBHs, \citealt{mapelli09}; \citealt{mapelli10}; \citealt{mapelli11a}; \citealt{mapelli11b}; M13). 

 Finally, we recall that the stellar evolution recipes adopted in this work (\citealt{hurley00}) are based on single-star evolution models, while recipes for binary evolution are the same as described in \citet{portegies01}. Recipes for binary evolution include prescriptions for angular momentum loss by magnetic stellar wind and/or by gravitational wave radiation, tidal circularization, circularization by gravitational wave emission, mass transfer and Roche lobe filling, common envelope, and binary merger.

\subsection{Initial conditions and simulation grid}\label{sec:sec2.1}
In this paper, we describe the results of 600 $N-$body simulations of young SCs (200 with metallicity $Z=0.01$ Z$_\odot$, 200 with  $Z=0.1$ Z$_\odot$, and the remaining 200 with  $Z=1$ Z$_\odot$). Each group of 200 simulations with the same metallicity is a different realization of the same SC model. We ran many different realizations of the same SC to obtain a sufficiently large sample of BH binaries and to damp stochastic fluctuations (each SC hosts $\sim{}8-9$ BHs on average, see M13). Half of the simulations presented in this paper are the same as described in M13. The remaining simulations are new runs. % (to increase the statistics).

 Each simulated SC is modelled as a spherical King profile with central dimensionless potential $W_0=5$ (\citealt{king66}) and $N_\ast{}=5500$ stars (corresponding to an initial total mass $M_{\rm TOT}\sim{}3000-4000$ M$_\odot{}$). %The resulting core density, at the beginning of the simulation, is $\rho{}_{\rm c}\sim{}2\times{}10^3$ M$_\odot{}$ pc$^{-3}$. 
%The main parameters adopted for the initial conditions are reported in 
Table~\ref{tab:tab1} shows the main initial properties of the simulated SCs. 
%%%%%%%%%%%%%%%%%%%%%%%%%%%%%%%%%%% FIGURE 1 %%%%%%%%%%%%%%%%%%%%%%%%%%%%%%%%%%
\begin{figure}
\center{{
\epsfig{figure=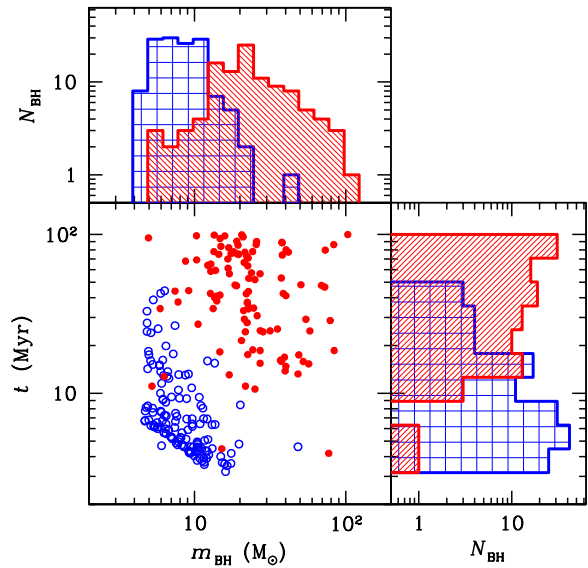,width=8.5cm}  %was exch_timevsMBH.eps
}}
\caption{\label{fig:fig1}
Main window: mass of the BHs that power a simulated RLO system versus the time when the RLO phase starts. Red filled circles: RLO-EBs (i.e. EBs that undergo RLO). Blue open circles: RLO-PBs (i.e. PBs that undergo RLO). The marginal histograms show the distribution of the masses of BHs that power RLO systems and the distribution of the times when the RLO phase starts ($t=0$ is the beginning of the simulation).  In both marginal histograms, the red hatched histogram (blue cross-hatched histogram) refers to  RLO-EBs (RLO-PBs). In this plot, all the simulated SCs are shown, and we do not distinguish between different metallicities.}
\end{figure}
%%%%%%%%%%%%%%%%%%%%%%%%%%%%%%%%%%%%%%%%%%%%%%%%%%%%%%%%%%%%%%%%%%%%%%%%%%%%%%

A primordial binary (PB) fraction $f_{\rm PB}=0.1$ was adopted.  This means that $\sim{}18$ per cent of all stars in the simulated SCs are members of a binary system, since $f_{\rm PB}$ is defined as the number of PBs in each SC divided by the number of `centers of mass' (CMs) in the SC (see Table~\ref{tab:tab1} for details). %$f_{\rm PB}$ is a delicate ingredient of our simulations. Recent observations show that the observed binary fraction can be very high in open and young SCs ($\gtrsim{}30\,{}$ per cent, with a large uncertainty, e.g. \citealt{sollima10}; \citealt{li13}). On the other hand, PBs are a bottleneck for direct-summation N-body codes. Thus, most direct-summation N-body simulations do not include PBs (or include a low fraction of PBs). In a forthcoming study, we will focus on the impact of PBs on the formation of X-ray binaries and other exotic binaries. In this paper, we briefly discuss the effects of a higher PB fraction in Appendix~A.

 The mass of single stars was randomly drawn according to a Kroupa initial mass function (IMF, \citealt{kroupa01}), with minimum mass $m_{\rm min}=0.1$ M$_\odot{}$ and maximum mass $m_{\rm max}=150$ M$_\odot{}$. The same procedure was adopted to select the mass of the primary member of a binary (i.e. the most massive member of a binary), while
% We make the further conservative assumption that the initial stellar distribution is not mass segregated (see e.g. Chatterjee et al. 2009 for the importance of this assumption).}. 
the mass of the secondary member ($m_2$) was drawn from a uniform distribution between $0.1\,{}m_1$  and $m_1$ (where $m_1$ is the mass of the primary). The initial semi-major axis $a$ of a binary was chosen from a distribution $f(a)\propto{}1/a$ (\citealt{sigurdsson95}; \citealt{portegies96}).  The minimum and the maximum allowed value of the semi-major axis are 1~R$_\odot{}$  and $10^5\,{}$R$_\odot{}$, respectively, but we discard systems whose periapsis is smaller than the sum of the radii of the two stars (\citealt{portegies07}). The maximum value of the semi-major axis is sufficiently large to include a number of soft binaries. 
%A different choice of the maximum value of $a$ in the range $\sim{}10^3-10^6$R$_\odot$ does not affect the results significantly (e.g.  Portegies Zwart et al. 2007). 
The initial eccentricity $e$ of a binary is randomly selected from a thermal distribution $f(e)=2\,{}e$, in the $0-1$ range (\citealt{heggie75}).

%The simulated SCs have half-mass relaxation time $t_{\rm hm}\sim{}10\,{}{\rm Myr}\,{}(r_{\rm hm}/0.8\,{}{\rm pc})^{3/2}\,(M_{\rm TOT}/3500\,{}{\rm M_\odot})^{1/2}$, where $r_{\rm hm}$ is the initial half-mass radius of the SC (in our simulations $r_{\rm hm}\sim{}0.8-0.9$ pc). Thus, the core collapse time (Portegies Zwart \&{} McMillan 2002) is $t_{\rm cc}\approx{}2\,{}{\rm Myr}\,{}(t_{\rm hm}/10\,{}{\rm Myr})$. 
We integrate the evolution of these SCs for the first 100 Myr. %, therefore studying a phase of the life of the cluster in which dynamical interactions are particularly intense. 
We recall that the half-mass relaxation timescale for these SCs is $t_{\rm rlx}\sim{}10\,{}{\rm Myr}\,{}(r_{\rm hm}/0.8\,{}{\rm pc})^{3/2}\,(M_{\rm TOT}/3500\,{}{\rm M_\odot})^{1/2}$, where $r_{\rm hm}$ is the initial half-mass radius of the SC (in our simulations $r_{\rm hm}\sim{}0.8-0.9$ pc). Thus, the core collapse timescale (\citealt{portegies02}) is $t_{\rm cc}\approx{}3\,{}{\rm Myr}\,{}(t_{\rm rlx}/10\,{}{\rm Myr})$. 

%We make three sets of runs corresponding to three different metallicities: 0.01 Z$_\odot{}$, 0.1 Z$_\odot{}$ and 1 Z$_\odot{}$. For each of these metallicities we simulate 100 different clusters, for a total of 300 SCs.
%The main parameters adopted for the initial conditions are reported in Table~1.

 The properties of the simulated SCs ($M_{\rm TOT}$, $r_{\rm c}$ and $r_{\rm hm}$) match those of observed dense young SCs (e.g. the Orion Nebula Cluster,  \citealt{portegies10}; see also \citealt{hillenbrand98}; \citealt{dias02}; \citealt{pfalzner09}; \citealt{kuhn12}).  Finally, our simulations do not account for the tidal field of the host galaxy.  The effect of tidal fields will be added and discussed in forthcoming papers. In Section~\ref{sec:caveat}, we discuss some additional caveats concerning our choice of the initial conditions.

 %%%%%%%%%%%%%%%%%%%%%%%%%%%%%%% TABLE 1%%%%%%%%%%%%%%%%%%%%%%%%%%%%%%%%%
\begin{deluxetable}{ll}
\tabletypesize{\tiny}
\tablewidth{0pt}
\tablecaption{Most relevant initial conditions.} %\leavevmode
\tablehead{
\colhead{Parameter} 
& \colhead{Values}} 
\startdata
$W_0$ & 5 \\
$N_\ast{}$ & 5500 \\
$r_{\rm c}$ (pc) & 0.4\\
$c$ & 1.03\\
IMF & Kroupa (2001)\\
$m_{\rm min}$ (M$_\odot{}$) & 0.1\\
$m_{\rm max}$ (M$_\odot{}$) & 150\\
$Z\,{}({\rm Z}_\odot{})$ & 0.01, 0.1, 1.0\\
$f_{\rm PB}$ & 0.1
%\noalign{\vspace{0.1cm}}
%\hline
%\end{tabular}
\enddata
%\begin{flushleft}
\tablecomments{
%\footnotesize{
{\it Notes.} $W_0$: central dimensionless potential in the \citet{king66} model; $N_{\ast}$: number of stars per SC;  $r_{\rm c}$: initial core radius; $c\equiv{}\log{}_{10}{(r_{\rm t}/r_{\rm c})}$: concentration ($r_{\rm t}$ is the initial tidal radius); IMF: initial mass function; $m_{\rm min}$ and $m_{\rm max}$: minimum and maximum simulated stellar mass, respectively; $Z$: metallicity of the SC; $f_{\rm PB}$: fraction of PBs,  defined as the number of PBs in each SC divided by the number of `centers of mass' (CMs) in the SC. In each simulated SC, there are initially 5000 CMs, among which 500 are designated as `binaries' and 4500 are `single stars' (see \citealt{downing10} for a description of this formalism). Thus, 1000 stars per SC are initially in binaries.\label{tab:tab1}}
%\end{flushleft}
%\end{center}
%\end{table}
\end{deluxetable}
%%%%%%%%%%%%%%%%%%%%%%%%%%%%%%%%%%%%%%%%%%%%%%%%%%%%%%%%%%%%%%%%%%%%%%%%%%%%%

%hange, this binary might be sufficiently tight to start RLO immediately after th

\section{Results}\label{sec:sec3}
\subsection{Formation pathways of BHs}\label{sec:sec3.1}
 Tables~\ref{tab:tab2} and \ref{tab:tab3} summarize the different formation pathways of simulated BHs, in the case of light BHs (LBHs, i.e. BHs with $m_{\rm BH}<25$ M$_\odot$) and  MSBHs (with $m_{\rm BH}\ge{}25$ M$_\odot$), respectively.
%%%%%%%%%%%%%%%%%%%%%%%%%%%%%%% TABLE 2%%%%%%%%%%%%%%%%%%%%%%%%%%%%%%%%%
\begin{deluxetable}{llllll}
\tabletypesize{\tiny}
 \tablewidth{0pt}
\tablecaption{Summary of the formation pathways for simulated LBHs (i.e. only the BHs with $m_{\rm BH}<25$ M$_\odot$).} %\leavevmode
%\begin{tabular}[!h]{llllll}
%\hline
\tablehead{\colhead{$Z$} & \colhead{Single star} & \colhead{Star in PB} & \colhead{Star-star merger}     & \colhead{BH-star merger}\\
 \colhead{(Z$_\odot$)} & \colhead{} & \colhead{} &  \colhead{}    & \colhead{}}      
\startdata %\hline
0.01 & 1310 (11) & 127   (57)      & 118   (0)        & 15 (0) \\ %from totlbh_single_W5_N5000_Z001.txt totlbh_binary_W5_N5000_Z001.txtb totlbh_binary_W5_N5000_Z001.txtm totlbh_truemerger_W5_N5000_Z001.txt
0.1  & 1318 (17) & 110 (45)        & 106   (2)        & 22 (0) \\
1.0  & 1432 (29) & 98  (38)        & 131   (2)        & 30 (1) \\
All$^{\rm a}$  & 4060 (57) & 335 (140) &  355 (4)         & 67 (1)
\enddata
\tablecomments{{\it Notes.} Column 1: metallicity of the SC; column 2: number of simulated LBHs that are born from single stars;  column 3: number of simulated LBHs that are born from stars in PBs (in column 3 we consider only the cases in which the PB does not merge before the formation of the BH; the cases in which the binary merges before the formation of the BH are listed in column 4);  column 4: number of simulated LBHs that are born from the merger of two stars (in $\sim{}95$ per cent of cases the two merging stars are two members of a PB); column 5: number of simulated LBHs that are born from the merger of a star and a BH.\\
The numbers within parentheses (in each column) refer to the number of LBHs that power RLO systems belonging to each formation pathway.  \\
$^{\rm a}$`All' refers to the statistics of all runs without distinguishing between different metallicities.\label{tab:tab2}}
%\end{flushleft}
%\end{center}
\end{deluxetable}
%%%%%%%%%%%%%%%%%%%%%%%%%%%%%%%%%%%%%%%%%%%%%%%%%%%%%%%%%%%%%%%%%%%%%%%%%%%%%
%%%%%%%%%%%%%%%%%%%%%%%%%%%%%%% TABLE 3%%%%%%%%%%%%%%%%%%%%%%%%%%%%%%%%%
\begin{deluxetable}{llllll}
\tabletypesize{\tiny}
 \tablewidth{0pt}
\tablecaption{Summary of the formation pathways for simulated MSBHs (i.e. only the BHs with $m_{\rm BH}\ge{}25$ M$_\odot$).} %\leavevmode
%\begin{tabular}[!h]{llllll}
%\hline
%\tablehead{\colhead{$Z$ (Z$_\odot$)} & \colhead{Single star} & \colhead{Star in PB} & \colhead{Star-star merger}     & \colhead{BH-star merger}}
\tablehead{\colhead{$Z$} & \colhead{Single star} & \colhead{Star in PB} & \colhead{Star-star merger}     & \colhead{BH-star merger}\\
\colhead{(Z$_\odot$)} & \colhead{} & \colhead{} &  \colhead{}    & \colhead{}}
\startdata %\hline
0.01 & 218 (19)  & 18 (1)          & 17 (0)         & 42 (1) \\ 
0.1  & 194 (13)  & 8 (0)          & 11  (1)         & 40 (4) \\
1.0  & 0   (0)  & 0 (0)          & 10   (1)         & 27 (2) \\
All$^{\rm a}$  & 412 (32) & 26 (1)   &  38  (2)        & 109 (7) 
\enddata
\tablecomments{{\it Notes.} The same as Table~\ref{tab:tab2} but for MSBHs.\label{tab:tab3}}
%\end{flushleft}
%\end{center}
\end{deluxetable}
%%%%%%%%%%%%%%%%%%%%%%%%%%%%%%%%%%%%%%%%%%%%%%%%%%%%%%%%%%%%%%%%%%%%%%%%%%%%%

Most LBHs come from single stars ($\sim{}85$ per cent), regardless of metallicity. Only $\sim{}6-8$ per cent of LBHs originate from stars in PBs, and $\sim{}7-8$ per cent form from the merger of two stars (the mergers occur mainly between two members of a PB). This is statistically consistent with the fact that $\sim{}18$ per cent of stars are members of PBs in our simulations. Finally, $\sim{}1-2$ per cent of LBHs come from the merger of a BH and a star. In the vast majority of cases the BH and the star were already members of a binary before merging, and most of these BH-star binaries were not PBs. In several cases, the formation of an unstable triple system triggers the BH-star merger (e.g.~\citealt{leigh13}). 

The numbers within parentheses in Table~\ref{tab:tab2} refer to the number of LBHs that power RLO systems belonging to each formation pathway. Remarkably, the majority of LBHs that power RLO systems come from PB members ($\sim{}54-84$ per cent, depending on metallicity), even if only  $\sim{}6-8$ per cent of LBHs originate from stars in PBs.

Most MSBHs at low metallicity ($Z=0.01$, 0.1 Z$_\odot$) come from the direct collapse of single stars ($\sim{}74-77$ per cent), while no MSBHs form from single stars at high metallicity ($Z=1$ Z$_\odot$), as a direct consequence of our stellar evolution models. 

The percentage of MSBHs that form from stars in PBs is very low (0, 3 and 6 per cent at Z$_\odot$, 0.1 Z$_\odot$ and 0.01 Z$_\odot$, respectively). The reason is that mass-transfer and common-envelope phases in tight PBs cause either the PB to merge or the first BH to be much lighter than in case of a single-star progenitor (as already shown by \citealt{linden10}).

The percentage of MSBHs that form from the merger of two stars is very low at low metallicity  (4 and 6 per cent at 0.1 Z$_\odot$ and 0.01 Z$_\odot$, respectively), while it is relevant at high metallicity (27 per cent at Z$_\odot$). Finally, the percentage of MSBHs that form from the merger of a BH and a star is non-negligible at low metallicity (16 and 14 per cent at  0.1 Z$_\odot$ and 0.01 Z$_\odot$, respectively) and is very large at high metallicity (73 per cent at Z$_\odot$). As for the LBHs, in the vast majority of cases the BH and the star were already members of a binary before merging, and most of these BH-star binaries were not PBs. We recall that the mass of the product of a merger between two stars or a star and a BH is likely overestimated in our simulations, because no mass loss is assumed during the encounter. In the case of a merger between two stars, mass loss by stellar winds is accounted for, after the merger.

The vast majority of MSBHs that power RLO systems form from single stars at low metallicity (72 and 91 per cent at 0.1 Z$_\odot$ and 0.01 Z$_\odot$, respectively). This implies that dynamical exchanges are very important for RLO systems powered by MSBHs (as we will discuss in the next sections). A very small fraction of   MSBHs that power RLO systems form from stars in PBs (only one system through all our simulations). 

At low metallicity (0.1, 0.01 Z$_\odot$), the number of MSBHs that  power RLO systems and form from the merger of two stars is negligible. In contrast, 33 per cent of MSBHs that  power RLO systems form from the merger of two stars at $Z={\rm Z}_\odot$. 

At $Z=0.01$ Z$_\odot$ the percentage of MSBHs that  power RLO systems and form from the merger of a star and a BH is negligible, while 22 per cent and 67 per cent of MSBHs powering RLO systems originate from the merger of a BH and a star at 0.1 Z$_\odot$ and Z$_\odot$, respectively. On the other hand, the absolute number of MSBHs that  power RLO systems and originate from the merger of a BH and a star is low.

\subsection{The importance of exchanges}\label{sec:sec3.2}
%%%%%%%%%%%%%%%%%%%%%%%%%%%%%%%%%%% FIGURE 2 %%%%%%%%%%%%%%%%%%%%%%%%%%%%%%%%%%
\begin{figure*}
\center{{
\epsfig{figure=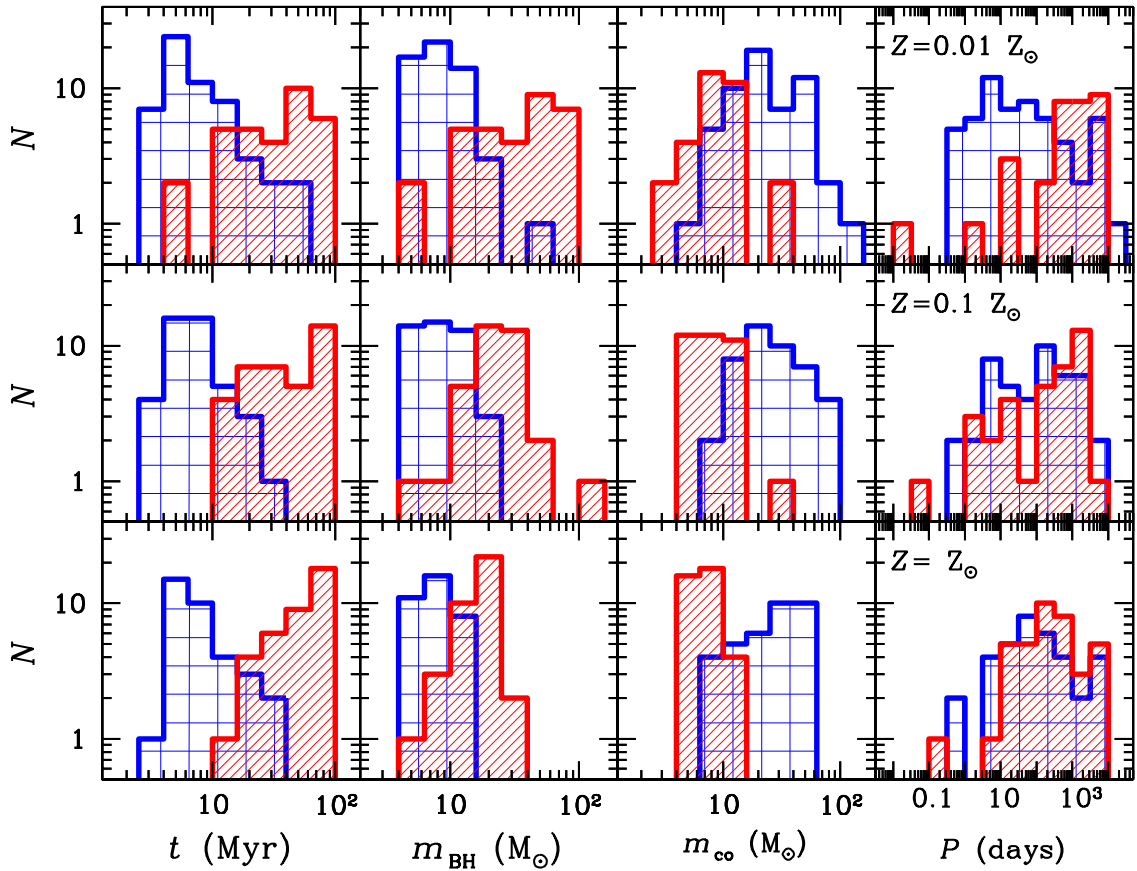,width=15.5cm} %was exchange_all.eps
}}
\caption{\label{fig:fig2}
From left to right: time $t$ when the RLO phase starts ($t=0$ is the beginning of the simulation), mass of the BH in RLO systems ($m_{\rm BH}$), mass of the donor star in RLO systems ($m_{\rm co}$), orbital period ($P$) at the beginning of the RLO phase. From top to bottom: $Z=0.01$ Z$_\odot$, 0.1  Z$_\odot$ and 1  Z$_\odot$. Red hatched histogram (blue cross-hatched histogram) refers to  RLO-EBs (RLO-PBs).}
\end{figure*}
%%%%%%%%%%%%%%%%%%%%%%%%%%%%%%%%%%%%%%%%%%%%%%%%%%%%%%%%%%%%%%%%%%%%%%%%%%%%%%

The best way to quantify the impact of SC dynamics on the formation of RLO systems is to distinguish between primordial binaries (PBs, i.e. binaries that were already in the initial conditions) and  binaries that formed through a dynamical exchange (EBs, i.e. binaries in which at least one of the two members entered the binary after a dynamical exchange). In our simulations (see Table~\ref{tab:tab4}), about $44$ per cent of RLO binaries are EBs (corresponding to 107 systems), while the remaining 56 per cent are PBs (corresponding to 137 systems). 
%The percentage of PBs that undergo RLO (hereafter RLO-PBs) is slightly different at different metallicity: ..,.. and .. at $Z=0.01$, 0.1 and 1 Z$_\odot{}$, respectively.

Fig.~\ref{fig:fig1} compares the behavior of PBs that undergo RLO (hereafter RLO-PBs, blue open circles) with that of EBs that undergo RLO (hereafter RLO-EBs, red filled circles). %The time when the RLO phase starts and the mass of the BH that powers the system are shown. 
RLO-EBs and RLO-PBs define two different groups: the former host typically more massive BHs than the latter, and tend to start the RLO phase later. The fact that RLO-EBs host more massive BHs is explained by the properties of exchanges: the probability for an exchange to occur is higher if the mass of the single object is higher than the mass of one of the former members of the binary (\citealt{hills89}; \citealt{hills92}).

 The fact that PBs tend to start the RLO phase at earlier times is an indication that RLO in these systems is mainly driven by stellar evolution rather than by dynamics. The majority of RLO-PBs start the accretion episode immediately after the formation of the first BH (3-7 Myr), when the companion star evolves towards the terminal-age MS (TAMS) and its radius increases rapidly, filling the Roche lobe (RL). In contrast, more time is needed for an EB to enter the RLO. In particular, most EBs form during the core collapse (which starts at $\sim{}3$ Myr and during which the maximum core density is reached, see \citealt{mapellibressan13}) and keep hardening (i.e. reducing their semi-major axis as a consequence of three-body encounters) for the next Myrs.

In Fig.~\ref{fig:fig1}, RLO systems from all simulated SCs are shown, without distinguishing between different metallicities. In Table~\ref{tab:tab4}, we consider different metallicities separately. %The number of simulated binaries that undergo RLO is 89, 82 and 73 at $Z=0.01$, 0.1 and 1 Z$_\odot{}$, respectively. The number of RLO-PBs (RLO-EBs) is 57, 45 and 35 (32, 37 and 38)  at $Z=0.01$, 0.1 and 1 Z$_\odot{}$, respectively. Thus,
There is an excess of  RLO-PBs at low metallicity. The most likely explanation is that stellar radii are smaller at low metallicity, allowing a larger number of PBs to avoid merger before the formation of the first BH and to start RLO (see \citealt{linden10}).

Fig.~\ref{fig:fig2} shows the main properties of RLO-PBs and RLO-EBs at different metallicities. EBs tend to enter a RLO phase later than PBs, regardless of metallicity. Furthermore, RLO-EBs tend to host more massive BHs than RLO-PBs. On the other hand, the maximum mass of the BHs powering the  RLO systems strongly depends on metallicity: at $Z=1$, 0.1 and 0.01  Z$_\odot$, the most massive BHs in RLO-EBs (RLO-PBs) have a mass $\sim{}38$, $103$ and $84$ M$_\odot$ ($\sim{}15$, 20 and 48 M$_\odot$), respectively. The maximum mass of BHs in RLO-EBs at $Z=1$ and 0.1  Z$_\odot$ is higher than the maximum BH mass that can be achieved through single-star evolution for these metallicities. Such high masses are the effect of BH-star mergers (see  Section~\ref{sec:sec3.1}). %(which lead to the formation of a Thorne-Zytkov star and then to the collapse into a BH). 

From Table~\ref{tab:tab4} we can see that metal-poor SCs ($Z=0.01-0.1$ Z$_\odot$) host a significant percentage (22--24 per cent) of RLO binaries powered by MSBHs (hereafter RLO-MSBHs). Solar-metallicity SCs host a small fraction (4 per cent) of  RLO-MSBHs, which were born from star-star and BH-star merger. We emphasize that all RLO-MSBHs but one  are EBs. The importance of RLO-MSBHs will be further investigated in the next Section.

Another difference between RLO-EBs and RLO-PBs is the mass range of donor stars: RLO-EBs have less massive donor stars than RLO-PBs. This is due to the fact that EBs start the RLO phase later than PBs, when the turn-off (TO) mass is smaller. The maximum mass of donor stars in RLO-PBs is clearly affected by stellar winds: the maximum mass of a donor star is $\gtrsim{}100$ M$_\odot$ at $Z=0.01$ Z$_\odot$, where stellar winds are negligible, while it is $\lesssim{}60$ M$_\odot$ at $Z=$ Z$_\odot$. %Finally, there is no significant difference between RLO-EBs and RLO-PBs if we look at the orbital periods.
Finally, the orbital periods tend to be longer in RLO-EBs than in RLO-PBs, especially at very low metallicity ($Z=0.01$ Z$_\odot$). Thus, the former tend to start the RLO phase only when the companion star evolves off the MS. RLO systems with longer period host either more massive BHs  (with large RL) or post-MS donor stars (with large radius).

\subsection{MSBHs versus light BHs}\label{sec:sec3.3}
%%%%%%%%%%%%%%%%%%%%%%%%%%%%%%%%%%% FIGURE 3 %%%%%%%%%%%%%%%%%%%%%%%%%%%%%%%%%%
\begin{figure}
\center{{
\epsfig{figure=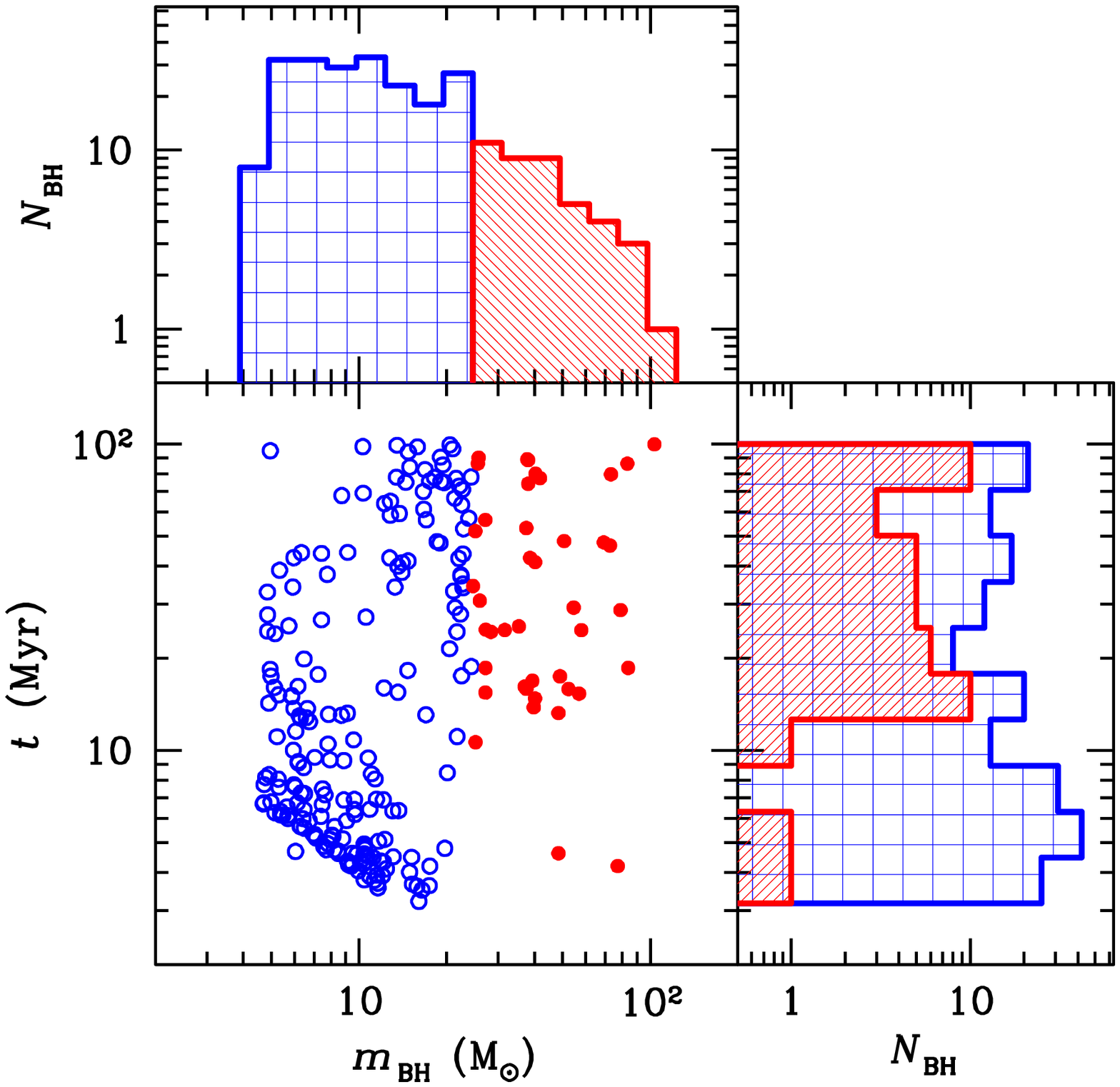,width=8.5cm} %was timevsMBH.eps
}}
\caption{\label{fig:fig3}
Main window: mass of the BHs that power simulated RLO systems versus the time when the RLO phase starts. Red filled circles: RLO-MSBHs (i.e. RLO systems powered by MSBHs); blue open circles: RLO-LBHs (i.e. RLO systems powered by BHs with mass $<25$ M$_\odot$). The marginal histograms show the distribution of the masses of BHs that power RLO systems and the distribution of the times when the RLO phase starts ($t=0$ is the beginning of the simulation).  In both marginal histograms, the red hatched histogram (blue cross-hatched histogram) refers to RLO-MSBHs  (RLO-LBHs). In this plot, as in Fig.~\ref{fig:fig1}, all the simulated SCs are shown, and we do not distinguish between different metallicities. }
\end{figure}
%%%%%%%%%%%%%%%%%%%%%%%%%%%%%%%%%%%%%%%%%%%%%%%%%%%%%%%%%%%%%%%%%%%%%%%%%%%%%%
%%%%%%%%%%%%%%%%%%%%%%%%%%%%%%%%%%% FIGURE 4 %%%%%%%%%%%%%%%%%%%%%%%%%%%%%%%%%%
\begin{figure*}
\center{{
\epsfig{figure=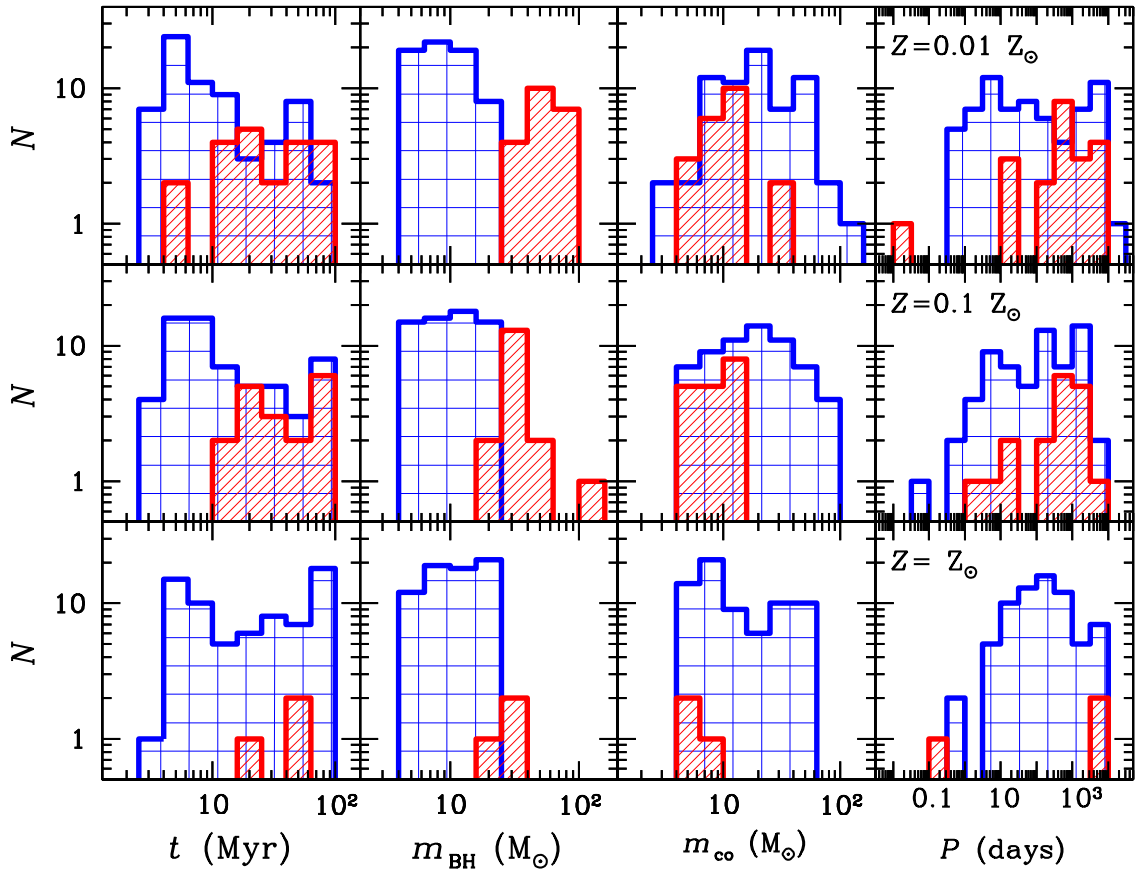,width=15.5cm} %was BH_all.eps
}}
\caption{\label{fig:fig4}
From left to right: time $t$ when the RLO phase starts ($t=0$ is the beginning of the simulation), mass of the BH in RLO systems ($m_{\rm BH}$), mass of the donor star in RLO systems ($m_{\rm co}$), orbital period at first RL approach ($P$). From top to bottom: $Z=0.01$ Z$_\odot$, 0.1  Z$_\odot$ and 1  Z$_\odot$. Red hatched histograms and blue cross-hatched histograms refer to  RLO-MSBHs and RLO-LBHs, respectively.}
\end{figure*}
%%%%%%%%%%%%%%%%%%%%%%%%%%%%%%%%%%%%%%%%%%%%%%%%%%%%%%%%%%%%%%%%%%%%%%%%%%%%%%%
%%%%%%%%%%%%%%%%%%%%%%%%%%%%%%% TABLE 4%%%%%%%%%%%%%%%%%%%%%%%%%%%%%%%%%
\begin{deluxetable}{llllll}
\tabletypesize{\tiny}
 \tablewidth{0pt}
\tablecaption{Summary of the properties of simulated RLO systems.} %\leavevmode
%\begin{tabular}[!h]{llllll}
%\hline
\tablehead{\colhead{$Z$} & \colhead{RLO} & \colhead{RLO-PBs}     & \colhead{RLO-EBs}      & \colhead{RLO-MSBHs}  & \colhead{RLO-LBHs}\\
\colhead{(Z$_\odot$)} & \colhead{} & \colhead{}     & \colhead{}      & \colhead{}  & \colhead{}}
\startdata %\hline
0.01 & 89  & 57          & 32           & 21         & 68 \\
0.1  & 82  & 45          & 37           & 18         & 64 \\
1.0  & 73  & 35          & 38           & 3          & 70 \\
All$^{\rm a}$  & 244 & 137         & 107          & 42         & 202 
%\noalign{\vspace{0.1cm}}
%\hline
%\end{tabular}
%\begin{flushleft}
%\footnotesize{
 \enddata
\tablecomments{{\it Notes.} $Z$ (column 1): metallicity of the SC; RLO (column 2): number of all simulated RLO systems;  RLO-PBs (column 3): number of PBs that undergo RLO;  RLO-EBs (column 4): number of EBs that undergo RLO; RLO-MSBHs (column 5): number of  RLO systems powered by MSBHs; RLO-LBHs (column 6): number of RLO systems powered by BHs with mass$<25$ M$_\odot$ for different metallicities and in total.\\
$^{\rm a}$`All' refers to the statistics of all runs without distinguishing between different metallicities.\label{tab:tab4}}
%\end{flushleft}
%\end{center}
\end{deluxetable}
%%%%%%%%%%%%%%%%%%%%%%%%%%%%%%%%%%%%%%%%%%%%%%%%%%%%%%%%%%%%%%%%%%%%%%%%%%%%%
As we mentioned in the previous section, RLO systems powered by MSBHs  (hereafter RLO-MSBHs) are basically a sub-group of RLO-EBs, since all RLO-MSBHs but one are EBs (see Figures~\ref{fig:fig1}, ~\ref{fig:fig2} and ~\ref{fig:fig3}). This confirms one of the most important results of M13, i.e. that dynamics is essential to drive the formation of RLO-MSBHs. As already shown by \citet{linden10}, isolated binaries can hardly evolve into RLO-MSBHs (even if they have a sufficiently small initial semi-major axis), because mass-transfer and common-envelope phases (before the formation of the first BH) cause either the two stars to merge or the first BH to be much lighter than in case of a single-star progenitor. Thus, RLO-MSBHs are expected to be very rare among isolated field binaries. In contrast, in a dense SC, single MSBHs can acquire companions through dynamical exchanges and power RLO systems.

Most simulated MSBHs are born from direct collapse of metal-poor stars ($\sim{}79$ per cent of all MSBHs in RLO systems),
%$\sim{}13$ per cent of all BHs are MSBHs at $Z=0.01-0.1$ Z$_\odot$, see Table~2 of Mapelli et al. 2013),
 but a fraction of MSBHs can form and/or grow in mass through mergers with other stars. This explains why there is a non-zero fraction of MSBHs even at solar metallicity. In particular, 9 MSBHs that power RLO systems (corresponding to $\sim{}21$ per cent of all RLO-MSBHs) come from the merger of either two stars or a BH and a star (Table~\ref{tab:tab3}). %Of these merger-born MSBHs, three are at $Z=$ Z$_\odot$, five at $Z=0.1$ Z$_\odot$ and one at  $Z=0.01$ Z$_\odot$.

In our simulations, 24 and 22 per cent of all BHs that power RLO systems are MSBHs at $Z=0.01$ and $0.1$ Z$_\odot$, respectively (see Table~\ref{tab:tab4}). Since  MSBHs are only $\sim{}18$ per cent of all BHs at $Z=0.01-0.1$ Z$_\odot$ (see Tables~\ref{tab:tab2} and \ref{tab:tab3}), this implies that MSBHs have a higher probability of powering RLO systems than low-mass BHs.  At $Z=$ Z$_\odot$ about 4 per cent of all BHs that power RLO systems are MSBHs born from mergers. 

Fig.~\ref{fig:fig4} compares the main properties of RLO-MSBHs  with those of RLO systems powered by `light' ($<25$ M$_\odot$) BHs (hereafter RLO-LBHs). RLO-MSBHs behave in the same way as other RLO-EBs: they start the RLO phase generally later than light BHs and have smaller donor stars.
%AGGIUNGERE TABELLA!!!

\subsection{Properties of the donor stars}\label{sec:sec3.3}
%Our $N-$body simulations include information about the time- and metallicity-dependent properties of the stars (mass, radius $R$, luminosity $L$ and effective temperature $T_{\rm eff}$). Thus, we can investigate the observational signatures of the donor stars at the time they fill their RL.
Our $N-$body simulations contain information about the mass, radius ($R$), luminosity ($L$) and effective temperature ($T_{\rm eff}$)  of the stars, and account for their dependence on time and metallicity. Thus, we can investigate the observational signatures of the donor stars at the time they fill their RL.

\begin{figure}
\center{{
\epsfig{figure=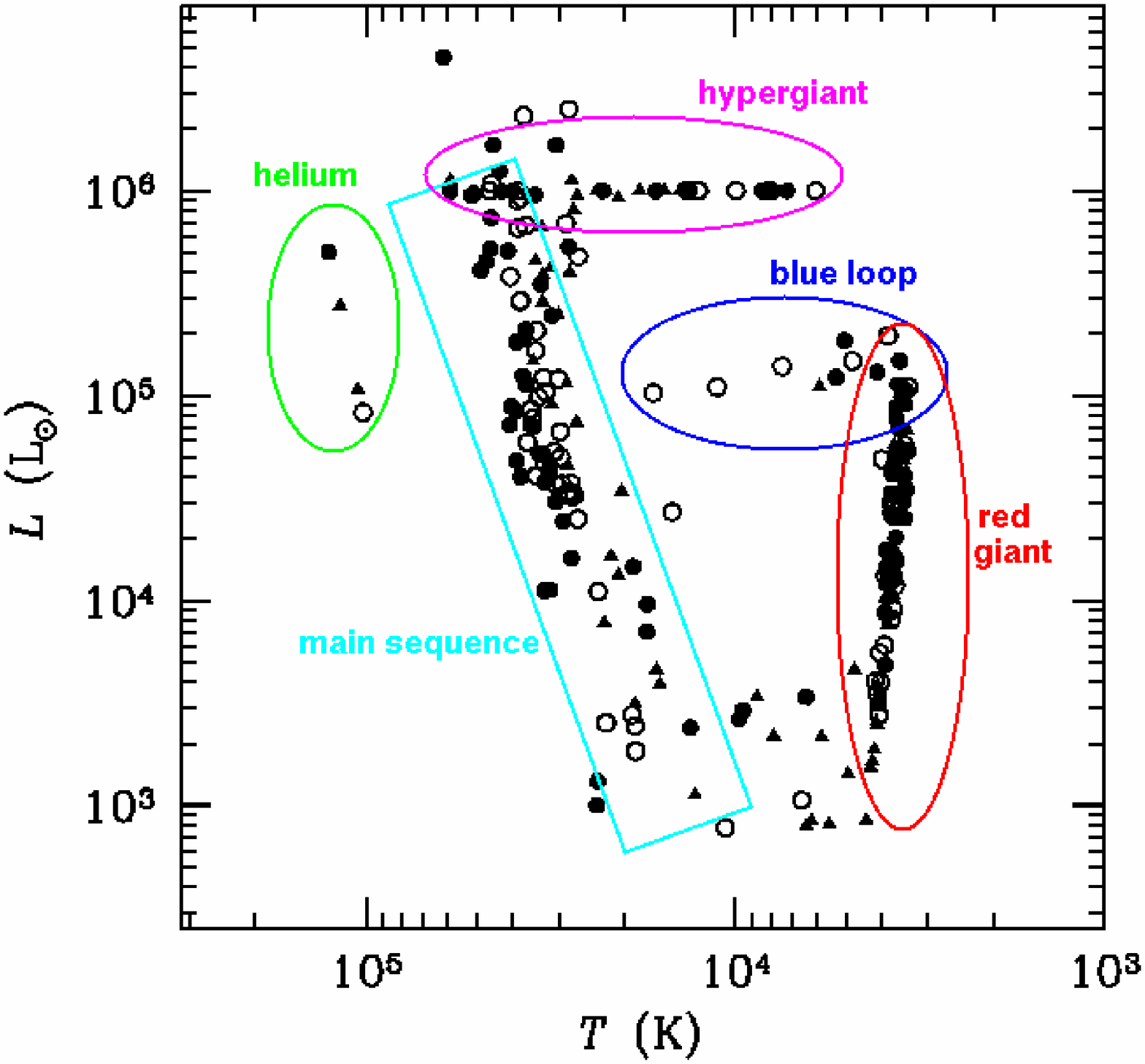,width=8.5cm} %was LeffTeff_ok_2.eps
}}
\caption{\label{fig:fig5}
HR diagram of the simulated donor stars when they fill the RL. Filled circles: $Z=0.01$ Z$_\odot$; open circles: $Z=0.1$ Z$_\odot$; filled triangles: $Z=$ Z$_\odot$. The colored ellipses and boxes approximately indicate the most common types of donor stars. In particular: MS (cyan box), red-giant and supergiant stars (red ellipse), blue loop stars (blue ellipse), hypergiant stars (magenta ellipse) and naked-core helium stars (green ellipse).}
\end{figure}

Fig.~\ref{fig:fig5} shows the Hertzsprung-Russell (HR) diagram of the donor stars when they fill the RL for the first time. It is apparent that a large portion of donor stars are in the MS or in the red-giant branch. Table~\ref{tab:tab5} summarizes the types of donor stars at the first RL approach. A large percentage of donor stars ($\sim{}33-50$ per cent, depending on the metallicity) are either in the MS or in the Hertzsprung gap (i.e. they just left the MS). A similar percentage ($28-40$ per cent) is either in the red-giant or in the red-supergiant branch.  A significant fraction ($15-21$ per cent) is composed of hypergiant stars.  Blue loop and  WR stars are much less frequent. Finally, two donor stars are in the LBV phase, corresponding to $\sim{}0.8$ per cent of all donor stars.

We notice that MS stars are the most important group of donor stars at low metallicity ($Z=0.01-0.1$ Z$_\odot$) while red-giant stars are more frequent at high metallicity ($Z=$ Z$_\odot$). The main reason is that the stellar radii  are smaller at low metallicity, %during the MS, 
allowing a larger number of PBs to avoid merger before the formation of the first BH and to start RLO when the donor is still a MS, or immediately after (see also the discussion in \citealt{linden10}). This  is supported by the fact that the number of RLO-PBs is $\sim{}40$ per cent higher  at $Z=0.01$ Z$_\odot{}$ than at $Z=1$ Z$_\odot{}$ (see Section~\ref{sec:sec3.2} and Table~\ref{tab:tab4}). 

Finally, we notice that a large portion of binaries that undergo RLO will merge at the end of the RLO phase or within the next 2 Myrs (76, 87 and 79 per cent at $Z=0.01$, 0.1 and 1 Z$_\odot{}$, respectively).
% from /MICMAP/star2ok/cineca_dataTOT/Z001/newanalysis/MDOT/newrun/ISTO/LZ/RL_tot_W5_N5000_Z1_paperII_LZ/RL_photometryandmergers_W5_N5000_Z1_paperII.txt

 %%%%%%%%%%%%%%%%%%%%%%%%%%%%%%% TABLE 5%%%%%%%%%%%%%%%%%%%%%%%%%%%%%%%%%
\begin{deluxetable}{lllllll}
%\begin{center}
\tabletypesize{\tiny}
 \tablewidth{0pt}
\tablecaption{Percentage of donor stars divided by type$^{\rm a}$.} %\leavevmode
%\begin{tabular}[!h]{llllll}
%\hline
\tablehead{\colhead{$Z$} & \colhead{MS \&   HG} & \colhead{Hy} & \colhead{RG} & \colhead{BL} & \colhead{WR} & \colhead{LBV}\\
\colhead{(Z$_\odot$)} & \colhead{(\%)} & \colhead{(\%)} & \colhead{(\%)} & \colhead{(\%)} & \colhead{(\%)} & \colhead{(\%)}} 
%\hline
\startdata
%MS (\%)                & 5, 17,  11      \\
0.01 & 50 & 15 & 31 & 1 & 2 & 1 \\
0.1  & 45 & 21 & 28 & 4 & 1 & 1 \\
1    & 33 & 19 & 40 & 5 & 3 & 0 \\
All$^{\rm b}$ & 43 & 19 & 33 & 3 & 2  & $<1$ %105.0  46.0 80.0 8.0 5.0
%\hline
%\noalign{\vspace{0.1cm}}
%\hline
%\end{tabular}
\enddata
%\begin{flushleft}
%\footnotesize
\tablecomments{{\it Notes.} $^{\rm a}$Percentages for each type of donor stars and for different metallicity. MS: main-sequence stars, HG: Hertzsprung gap stars (generally stars that just left the MS), Hy: hypergiant stars, RG: red-giant stars and red-supergiant stars, BL: blue loop stars, WR: Wolf-Rayet stars, LBV: luminous blue variable stars.\\
$^{\rm b}$`All' refers to the statistics of all runs without distinguishing between different metallicities.\label{tab:tab5}}
%\end{flushleft}
%\end{center}
\end{deluxetable}
%%%%%%%%%%%%%%%%%%%%%%%%%%%%%%%%%%%%%%%%%%%%%%%%%%%%%%%%%%%%%%%%%%%%%%%%%%%%%
 %%%%%%%%%%%%%%%%%%%%%%%%%%%%%%% TABLE 2%%%%%%%%%%%%%%%%%%%%%%%%%%%%%%%%%
%\begin{table}
%\begin{center}
%\caption{Percentage of donor stars divided by type.} \leavevmode
%\begin{tabular}[!h]{ll}
%\hline
%Donor type         & Percentage$^{\rm a}$ (\%)\\
%                   & at $Z=0.01$, 0.1, 1 Z$_\odot$\\
%\hline
%MS (\%)                & 5, 17,  11      \\
%MS and Hertzsprung gap                 & 50, 45,  33      \\
%%Hertzsprung gap (\%)   & 25, 28,  22   \\
%Hypergiant         & 16, 22,  19      \\
%Red-giant          & 31, 28,  40    \\
%Blue Loop          & 1,   4,  5     \\
%WR                 & 2,   1,  3 \\
%\hline
%\noalign{\vspace{0.1cm}}
%\hline
%\end{tabular}
%\begin{flushleft}
%\footnotesize{$^{\rm a}$In column 2, from left to right, the three different percentages for each type (separated by a comma)  refer to metallicity $Z=0.01$, 0.1 and 1 Z$_\odot$. The type 'Red-giant' star  refers to both red-giant stars and red-supergiant stars.}
%\end{flushleft}
%\end{center}
%\end{table}
%%%%%%%%%%%%%%%%%%%%%%%%%%%%%%%%%%%%%%%%%%%%%%%%%%%%%%%%%%%%%%%%%%%%%%%%%%%%%
%%%%%%%%%%%%%%%%%%%%%%%%%%%%%%%%%%% FIGURE 5 %%%%%%%%%%%%%%%%%%%%%%%%%%%%%%%%%%
%%%%%%%%%%%%%%%%%%%%%%%%%%%%%%%%%%%%%%%%%%%%%%%%%%%%%%%%%%%%%%%%%%%%%%%%%%%%%%

\subsection{The contribution of the accretion disk}
In the previous Section, we analyzed the observational signatures of the simulated donor stars. On the other hand, %the X-ray photons emitted from the inner parts of the accretion disk are likely reprocessed by the donor star and by the accretion disk itself, producing more optical photons. 
the contribution of the irradiated accretion disk might significantly affect the optical luminosity and colors of the observed counterparts (e.g. \citealt{copperwheat05}; \citealt{copperwheat07}; \citealt{mucciarelli07}; \citealt{patruno08}; \citealt{patruno10}; \citealt{madhusudhan08}; \citealt{grise12}). 
%\subsection{Ejections}

%%%%%%%%%%%%%%%%%%%%%%%%%%%%%%%%%%% FIGURE 6 %%%%%%%%%%%%%%%%%%%%%%%%%%%%%%%%%%
%\begin{figure}
%\center{{
%\epsfig{figure=fdisc_star,width=7.5cm} 
%}}
%\caption{\label{fig:fig6}
%Fraction of integrated flux coming from the irradiated disk ($F_{\rm disk}$) over the integrated flux coming from the donor star ($F_{\rm co}$) in the simulations {\bf LUCA: i flussi sono su tutto lo spettro o solo infrarosso-ottico?}. Blue horizontally-hatched histogram: $Z=1$ Z$_\odot$; black diagonal-hatched histogram: $Z=0.1$ Z$_\odot$; red vertically-hatched histogram: $Z=0.01$ Z$_\odot$.}
%\end{figure}
%%%%%%%%%%%%%%%%%%%%%%%%%%%%%%%%%%%%%%%%%%%%%%%%%%%%%%%%%%%%%%%%%%%%%%%%%%%%%%

Thus, we use the code described in Patruno \&{} Zampieri (2008, 2010) to model the optical emission associated with the simulated RLO binaries. The code accounts for both the emission of the donor star and the emission due  to X-ray reprocessing of the accretion disk and of the donor star. 
%to self-irradiation of the disk and to X-ray irradiation of the companion. A standard
A Shakura-Sunyaev disk is assumed. %and Eddington luminosity are assumed. 
If the accretion rate reaches the Eddington limit (for a standard accretion efficiency of $\sim{}0.1$), the luminosity is limited at Eddington and the excess mass is assumed to be expelled from the system. %A different treatment for the accretion is presently investigated
 A simplified description of radiative transfer for the interaction of
the X-rays with the disk and donor surfaces is adopted (an X-ray
illuminated plane-parallel atmosphere in radiative equilibrium; e.g. \citealt{copperwheat05}). The model does not include a Comptonization of disk emission in a corona. 
%{\bf Luca E gli altri parametri? inclination ? albedo? f\_irradiated ?}
We assume inclination $i=0$ (face-on disk) and albedo $=0.9$. The fraction of X-ray flux thermalized in the outer irradiated disk is typically in the range $0.004-0.008$.

The $N-$body simulations provide information about (i) the radius, mass, optical luminosity, effective temperature and age of the donor star when it fills its RL; (ii) the mass of the BH; (iii)  the orbital properties of the binary when it starts the RLO phase (orbital period). These properties of the simulated RLO systems are fed to the \citet{patruno08} code, to model the optical luminosity and colors.

The results indicate the overall importance of disk irradiation: the flux from the disk ($F_{\rm disk}$) ranges from $\sim{}0.1$ to $\sim{}10^4$ times the flux from the donor star ($F_{\rm co}$). The highest values of $F_{\rm disk}/F_{\rm co}$ are associated with the most massive MSBH systems ($Z=0.01$ Z$_\odot$), which have the most extended accretion disks and the faintest donor stars (red giant stars with $M\sim{}5-10$ M$_\odot{}$).

\begin{figure}
\center{{
\epsfig{figure=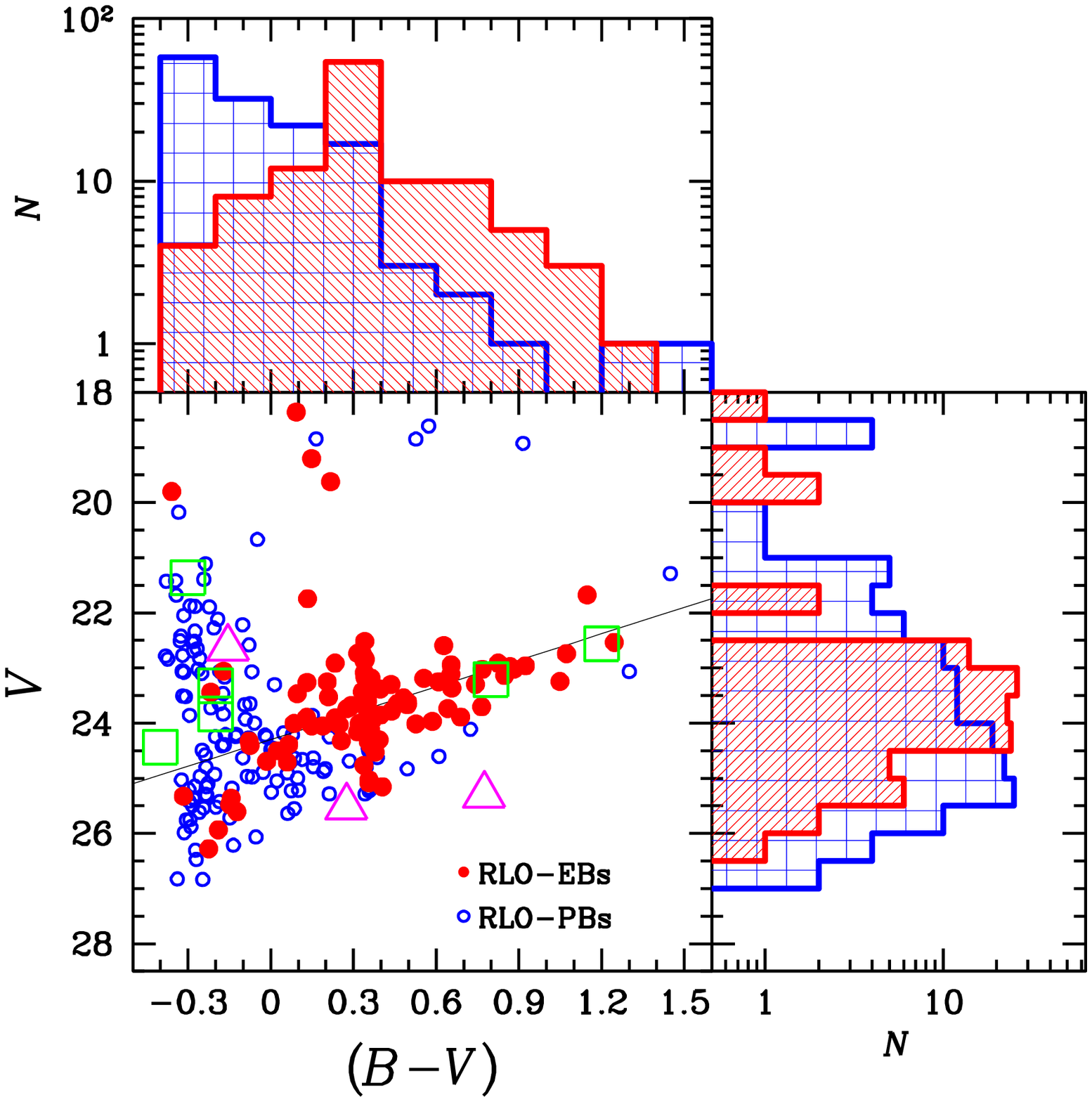,width=8.5cm} %was fig7.eps gladstone_only1st.eps
}}
\caption{\label{fig:fig6}
Main window: CMD (in the Johnson $B$ and $V$ filters) of the simulated RLO systems. The magnitudes take into account both the flux from the donor star and the flux from the irradiated disk (see the main text for details). Red filled circles: RLO-EBs (i.e. EBs undergoing RLO); blue open circles: RLO-PBs (i.e. PBs undergoing RLO). In the marginal histograms, red diagonal-hatched histogram:  RLO-EBs; blue cross-hatched histogram: RLO-PBs.
Green open squares:  counterparts of six ULXs from \citealt{gladstone13} (2013, M81 X-6, Ho~IX X-1, NGC~1313 X-1, NGC~1313 X-2, IC~342 X-1 and NGC~3034 ULX5) in the F435W and F555W {\it HST} filters. Magenta open triangles: counterparts of three ULXs from \citealt{gladstone13} (2013, NGC~2403 X-1, M83 XMM1 and NGC~5204 X-1) in the F435W and F606W {\it HST} filters (the F606W filter was adapted to the F555W filter as discussed in Appendix~\ref{appendix:counterparts}). Both the simulated and the observed magnitudes are apparent magnitudes (in Vega mag) obtained assuming that all sources are at 5 Mpc distance. The solid line corresponds to the $V\approx{}24.3-1.6\,{}(B-V)$ sequence of  RLO-EBs (see the text for details).}
\end{figure}

\begin{figure}
\center{{
\epsfig{figure=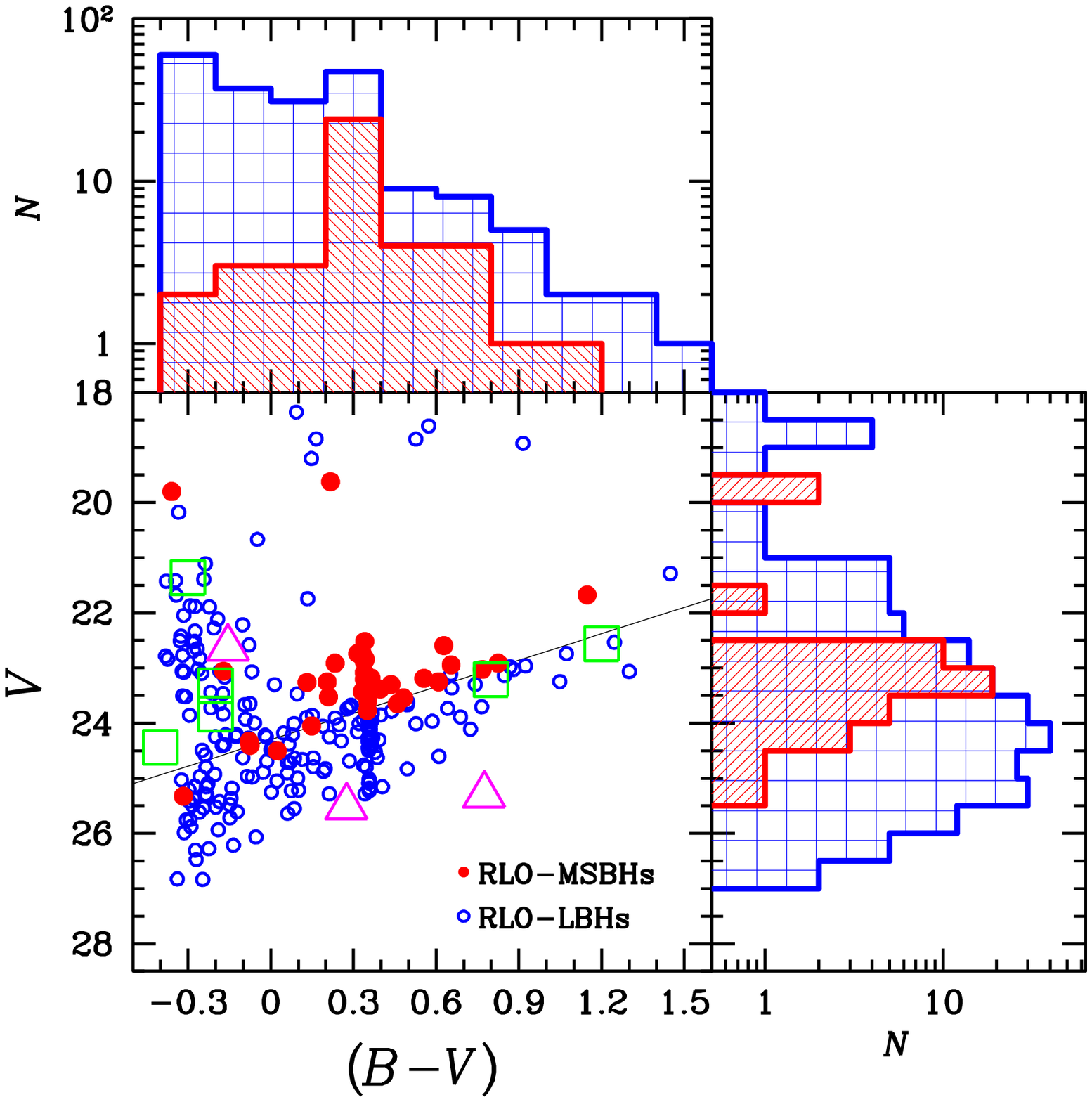,width=8.5cm} %was fig8.eps gladstone_only1st.eps
}}
\caption{\label{fig:fig7}
The same as Fig.~\ref{fig:fig6}, but for RLO-MSBHs and RLO-LBHs. In particular, in the main window, red filled circles: RLO-MSBHs; blue open circles: RLO-LBHs. In the marginal histograms, red diagonal-hatched histogram:  RLO-MSBHs; blue cross-hatched histogram: RLO-LBHs.}
\end{figure}

Figures~\ref{fig:fig6} and ~\ref{fig:fig7} show the $V-$band magnitude and the $B-V$ color (Johnson filters) of the simulated RLO systems, 
%adopting the $B$ and $V$ Johnson filters and
 assuming a distance of $5$ Mpc. RLO-EBs and  RLO-PBs are compared in  Fig.~\ref{fig:fig6}, while  RLO-MSBHs and  RLO-LBHs are compared in  Fig.~\ref{fig:fig7}.  
%%%%%%%%%%%%%%%%%%%%%%%%%%%%%%%%%%% FIGURE 6 %%%%%%%%%%%%%%%%%%%%%%%%%%%%%%%%%%
%%%%%%%%%%%%%%%%%%%%%%%%%%%%%%%%%%%%%%%%%%%%%%%%%%%%%%%%%%%%%%%%%%%%%%%%%%%%%%

%%%%%%%%%%%%%%%%%%%%%%%%%%%%%%%%%%% FIGURE 7 %%%%%%%%%%%%%%%%%%%%%%%%%%%%%%%%%%
%%%%%%%%%%%%%%%%%%%%%%%%%%%%%%%%%%%%%%%%%%%%%%%%%%%%%%%%%%%%%%%%%%%%%%%%%%%%%%

The $B-V$ color shows interesting features. The bluer RLO systems [$-0.35<(B-V)<0$] span a large range of magnitudes, from $V\sim{}20$ to $V\sim{}27$. Instead, the redder RLO systems define a narrower sequence in Figures~\ref{fig:fig6} and ~\ref{fig:fig7}, which can be approximated as $V\approx{}24.3\,{}-\,{}1.6\,{}(B-V)$ with a conspicuous scatter, larger at bluer colors\footnote{$V\approx{}24.3\,{}-\,{}1.6\,{}(B-V)$ comes from the least square fit of the simulated RLO-EBs, excluding the four outliers with $V<20.5$.}.

\begin{figure}
\center{{
\epsfig{figure=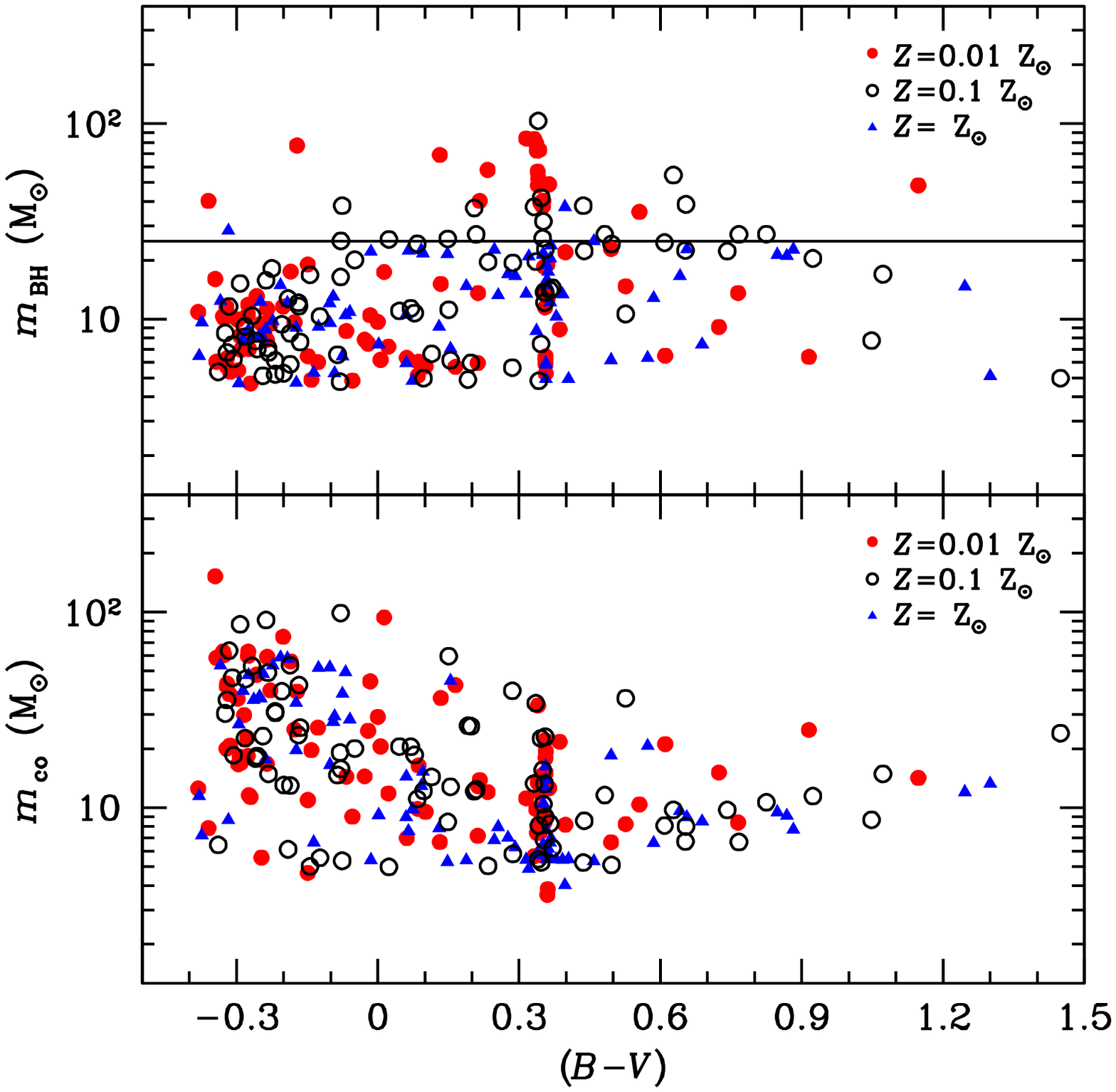,width=8.5cm} %was fig9.eps 
}}
\caption{\label{fig:fig8}
Top (bottom) panel: mass of the BH (donor star) of the simulated RLO systems versus the $B-V$ color. In both panels, filled red circles: systems with $Z=0.01$ Z$_\odot$, open black circles: systems with $Z=0.1$ Z$_\odot$,  filled blue triangles: systems with $Z=1$ Z$_\odot$. The horizontal line in the top panel marks the separation between light BHs and MSBHs (i.e. the threshold mass $m_{\rm BH}=25$ M$_\odot$). The simulated  magnitudes are apparent magnitudes (in Vega mag) obtained assuming that all sources are at 5 Mpc distance.}
\end{figure}

This behavior is explained by the mass range of the donor stars. The most massive young donor stars are generally associated with blue counterparts, while lower mass donor stars are found in both blue and red counterparts (depending on the relative importance of the disk). Thus, blue systems have a large scatter in $V$, since they can host both massive ($>30$ M$_\odot$), very bright donor stars and lower-mass (5--20 M$_\odot$), less luminous donor stars (see Fig.~\ref{fig:fig8}). In contrast, red systems are generally disk dominated and host only lower-mass (5--20 M$_\odot$) older stars, mostly red giant stars. %{\bf From Fig.~\ref{fig:fig8} it is also apparent that several RLO systems with $m_{\rm co}\sim{}5-20$ M$_\odot$ have a similar color $(B-V)\sim{}0.3-0.4$. This is due to the fact that the donor stars are red giant stars (red giant stars have very similar temperatures in our simulations, see Fig.~\ref{fig:fig5}), and disk irradiation is not important in these systems (thus the optical emission comes mainly from the donor star).}

Also, we notice that the sequence defined by  $V\approx{}24.3\,{}-\,{}1.6\,{}(B-V)$ is mainly populated by RLO-EBs (see Fig.~\ref{fig:fig6}), whose donor stars are predominantly older and less-massive than those of RLO-PBs (see Fig.~\ref{fig:fig2}). In contrast, RLO-PBs are mostly associated with the bluer counterparts. In both cases, there are some remarkable exceptions (i.e. very blue RLO-EBs and very red RLO-PBs).

Even in the CMD, RLO-MSBHs behave as a sub-group of RLO-EBs, since the former populate the same sequence as the latter. On the other hand,  RLO-MSBHs represent the high-luminosity tail of  RLO-EBs, since they populate mainly the upper envelope of the sequence. RLO-MSBHs span a narrower range of $V$ magnitudes ($19.5<V<25.5$) than RLO-LBHs, and $\sim{}80$ per cent of RLO-MSBHs  have $22<V<24$. The counterparts of RLO-MSBHs are predominantly red, with a peak at $B-V\sim{}0.3$.

 The position of the counterparts in the CMD of Figures~\ref{fig:fig6} and \ref{fig:fig7} can be understood as follows. 
The majority of the blue ($B-V<-0.1$) systems have MS companions, with masses from $\sim{}10$ M$_\odot$ to several tens of M$_\odot$. Very massive companions ($> 20$ M$_\odot$) are sufficiently bright that their intrinsically blue emission dominates over that of the accretion disk. Lower mass companions ($10-20$ M$_\odot$) are weaker and  the contribution of the disk is significant for them. They typically reside in tighter systems ($P<5-10$ d). As a consequence, their disk is not very extended and the combined disk plus donor emission appears as blue as that of higher mass systems.

On the other hand, donors in the $10-20$ M$_\odot$ range can also reside in systems with intermediate colors ($-0.1<B-V<0.2$), if they are slightly more evolved and at larger orbital separations ($P>5-10$ d). The disk is then more extended and its optical spectrum appears redder. %A minor fraction of MSBHs at low $Z$ are hosted in these systems. 
Markedly red counterparts ($B-V>0.2$) are produced by large-separation systems with very extended disks, the outer parts of which are strongly irradiated and emit significantly in the near-infrared band. For the largest systems ($P>100$ d) the flux of the $10-20$ M$_\odot$ evolved companion may overcome that of the disk. These systems host most MSBHs, since they typically form at later times (when $>20$ M$_\odot$ stars have already evolved) and from dynamical interactions. %(which tend to produce initially wider systems).

%MM: non ho capito se questo \`e in aggiunta o sostituzione di qualcosa che ho scritto io.}

%RLO-LBHs span a large range in $V$ band (18--27), but $\sim{}90$ ($\sim{}40$) per cent of systems have $22<V<26$ ($22<V<24$). RLO-MSBHs span a narrower range of $V$ magnitudes ($19.5<V<25.5$) and $\sim{}80$ per cent of systems have $22<V<24$.

%The $B-V$ color shows interesting features. The bluer RLO systems ($-0.35<(B-V)<0$) span a large range of magnitudes, from $V\sim{}20$ to $V\sim{}27$. Instead, the redder RLO systems define a narrower sequence, which can be approximated as $V\approx{}24.3\,{}-\,{}1.6\,{}(B-V)$.

%This behaviour is explained by the mass range of the donor stars. The most massive young donor stars are generally associated with blue counterparts, while lower mass donor stars are found in both blue and red counterparts (depending on the relative importance of the disk). Thus, blue systems have a large scatter in $V$ as they can host both massive ($>30$ M$_\odot$), very bright donor stars and lower-mass (10--20 M$_\odot$), less luminous donor stars (see Fig.~\ref{fig:fig7}). In contrast, red systems are generally disk dominated and host only lower-mass (10--20 M$_\odot$) older stars, mostly red giant stars.

%Furthermore, the mass of the BH colors contain a fingerprint of the mass of the BH....

Finally, the eight systems with $V<20$ are outliers: four of them are systems that are undergoing a merger, while in the other four systems the donor star evolves very rapidly  and the $N-$body outputs are not frequent enough to capture the fast changes of the properties of the donor star (i.e. there is a mismatch between the last time-step in which the radius was calculated and the time when the RLO phase starts).

%%%%%%%%%%%%%%%%%%%%%%%%%%%%%%%%%%% FIGURE 8 %%%%%%%%%%%%%%%%%%%%%%%%%%%%%%%%%%
%%%%%%%%%%%%%%%%%%%%%%%%%%%%%%%%%%%%%%%%%%%%%%%%%%%%%%%%%%%%%%%%%%%%%%%%%%%%%%

\section{Discussion}
\subsection{Ultraluminous X-ray sources and simulated RLO binaries}
Figures~\ref{fig:fig6} and \ref{fig:fig7} show that the simulated RLO systems describe a well-defined pattern in the CMD. In particular, RLO-EBs and RLO-PBs behave in two different ways, since the former are generally redder than the latter. In the CMD, RLO-MSBHs follow the same trend as RLO-EBs, since all  RLO-MSBHs but one  formed from dynamical exchange. On the other hand, RLO-MSBHs represent the high-luminosity envelope of RLO-EBs in the CMD.

%Furthermore, these differences translate into sin

These results provide important insights to understand the observations of bright X-ray binaries in young SCs. Comparing the simulations with a complete set of data is beyond the aims of this paper. In this section, we will focus on a specific class of X-ray binaries, the ULXs. We chose the ULXs because they are mostly associated with star forming regions and/or young SCs (e.g. \citealt{zezas02}; \citealt{kaaret04}; \citealt{berghea09}; \citealt{mapelli10}; \citealt{poutanen13}; \citealt{berghea13}), because they seem to anti-correlate with the local metallicity (e.g. \citealt{pakull02}; \citealt{zampieri04}; \citealt{soria05}; \citealt{swartz09}; \citealt{mapelli09}; \citealt{mapelli10}; \citealt{mapelli11a}; \citealt{kaaret13}; \citealt{prestwich13}) and because they have an estimated luminosity (assuming isotropy) higher than the Eddington luminosity of a 10 M$_\odot$ BH. For the high luminosity and for the anti-correlation with metallicity, \citet{mapelli09} proposed that (a fraction of) ULXs are associated with MSBHs (see also \citealt{zampieri09}; \citealt{mapelli10}; \citealt{mapelli11a}).

Unfortunately, identified counterparts are available in the literature only for a limited number of ULXs. The most recent and homogeneous sample of optical ULX counterparts has been reported by \citet{gladstone13} (2013, hereafter G13).  Among the 22 ULXs studied in G13 that have {\it Chandra} observations, and that have at least one detected possible counterpart from {\it Hubble Space Telescope} ({\it HST}) data, we selected nine sources (see Table~\ref{tab:tab6}), based on the criteria described in Appendix~\ref{appendix:counterparts}. 

 Five of the nine considered sources (Ho~IX X-1, NGC~1313 X-2, IC~342 X-1, NGC~5204 X-1 and NGC~3034 ULX5) have multiple possible counterparts in the {\it Chandra} error box. In Table~\ref{tab:tab6}, and in Figures~\ref{fig:fig6} and \ref{fig:fig7}, we report only the counterparts that were selected based on the criteria described in Appendix~\ref{appendix:counterparts}. In Figures~\ref{fig:fig6} and \ref{fig:fig7}, we show apparent magnitudes obtained positioning both the observed counterparts and the simulated ones at a distance of 5 Mpc. 
%assuming a distance of 5 Mpc for both the observed counterparts and the simulated ones. The observed counterparts were corrected for Galactic extinction.

%Of the 33 ULXs observed with {\it Hubble Space Telescope} ({\it HST}) and {\it Chandra} and studied in G13, nine have no visible counterparts. The remaining 22 ULXs have at least one detected possible counterpart. Among them, we selected nine ULXs (see Table~\ref{tab:tab6}), {\bf according to the criteria described in Appendix~\ref{appendix:counterparts}}. 
%13 have one possible optical counterpart, while multiple sources are in the error box of the other 9 ULXs. 

 %%%%%%%%%%%%%%%%%%%%%%%%%%%%%%% TABLE 6%%%%%%%%%%%%%%%%%%%%%%%%%%%%%%%%%
\begin{deluxetable*}{llllllllllll}
%\begin{center}
 \tabletypesize{\tiny}
 \tablewidth{0pt}
\tablecaption{Selected ULX counterparts from G13.} %\leavevmode
%\begin{tabular}[!h]{llllllllll}
%\hline
\tablehead{\colhead{Name}         & \colhead{ID}  & \colhead{DM}     &  \colhead{F435W}  & \colhead{F555W} & \colhead{F606W}    &   \colhead{$V$}  & \colhead{$(B-V)$} &  \colhead{$Z$} &  \colhead{$P$} & \colhead{$L_{\rm X,\,{}max}$} & \colhead{Ref.} \\
    \colhead{}     & \colhead{}  & \colhead{}   & \colhead{}   & \colhead{} & \colhead{}    &  \colhead{}   & \colhead{} &   \colhead{(Z$_\odot$)} & \colhead{(days)} & \colhead{($10^{39}$ erg s$^{-1}$)} & \colhead{}}
%\hline                                                                
\startdata
M81 X-6      & 1   & 31.13  &  23.7	& 24.0  & 23.8   &   21.36 & -0.3  & $0.40-0.22$ & $-$ & 6.7 & (1) \\ 
Ho~IX X-1     & 1   & 27.67  &  22.3	& 22.5	& NO     &   23.32 &  -0.2 & 0.38 & $-$ & 28  & (2) \\
%Ho~IX X-1     & 2   & 27.67  &  ND	& ND	& NO     &   --    & --    & 28 \\
%Ho~IX X-1     & 3   & 27.67  &  26.0	& 26.0	& NO     &   26.82 & 0.0   & 28 \vspace{0.2cm}\\
NGC~1313 X-1  & 1   & 28.06  &  23.6	& 24.0	& 23.7   &   24.43 & -0.4  & 0.10 & $-$ & 15  & (3) \\
NGC~1313 X-2  & 1   & 28.06  &  23.2	& 23.4	& NO     &   23.83 & -0.2  & 0.10 & $6,\,{}12$ & 8.1 & (3)\\
%NGC~1313 X-2  & 2   & 28.06  &  ND       & 24.1  & NO     &   24.53 & --    & 8.1\vspace{0.2cm}\\
IC~342 X-1    & 1   & 27.58  &  23.1	& 22.3	& 21.9   &   23.21 & 0.8   & $0.37-0.11$ & $-$ &  13.3 & (2) \\
%IC~342 X-1    & 2   & 27.58  &  25.0	& 24.0	& 24.0   &   24.91 & 1.0   & 13.3\vspace{0.2cm}\\
NGC~2403 X-1  & 1   & 27.70  &  25.0	& NO	& 24.5   &   25.52 & 0.275 & $0.29-0.18$ & $-$ & 3.6  & (1) \\
M83 XMM1     & 1   & 28.41  &  26.0	& NO	& 25.0   &   25.31 & 0.775 & $0.42-0.32$ & $-$ & 2.8  & (1) \\
NGC~5204 X-1  & 1   & 28.38  &  22.37 	& NO	& 22.3   &   22.64 & -0.155 & $-$ & $-$ & 8.7   & (2) \\
%NGC~5204 X-1  & 2   & 28.38  &  20.8	& NO	& 20.0   &   20.34 & 0.575 & 8.7\vspace{0.2cm}\\
NGC~3034 ULX5 & 1   & 27.73  &  23.0	& 21.8	& NO     &   22.56 & 1.2   & $-$ & 62 & 79    & (4) 
%NGC~3034 ULX5 & 2   & 27.73  &  21.9     & ND    & NO     &   --    & --    & 17.0\\
%\hline
%\noalign{\vspace{0.1cm}}
%\hline
%\end{tabular}
\enddata
%\begin{flushleft}
%\footnotesize{
\tablecomments{{\it Notes.} Column 1: ULX name; column 2: identification number of the counterpart from Table~4 of G13; column 3: distance modulus (DM); column 4: magnitude in F435W filter (from Table~4 of G13); column 5: magnitude in F555W filter (from Table~4 of G13); column 6: magnitude in F606W filter (from Table~4 of G13); column 7: $V$ magnitude obtained positioning the source at 5 Mpc. $V$ is obtained from F555W or, when F555W is not available, from F606W (converted into F555W as described in the text). $V$ is shown in Figures~\ref{fig:fig6} and ~\ref{fig:fig7}. Column 8: $(B-V)$ color. The $B$ magnitude is obtained from the F435W filter, positioning the source at 5 Mpc. $V$ is the same as tabulated in column 7. $(B-V)$ is shown in Figures~\ref{fig:fig6} and ~\ref{fig:fig7}.  Column 9: Metallicity $Z$ of the host galaxy. The integrated metallicity of the galaxy is given for Ho~IX and NGC~1313, while the metallicity range indicated for the other galaxies corresponds to the metallicity between 0.4 and $1\,{}R_{25}$ ($R_{25}$ being the isophotal radius, \citealt{devaucouleurs91}). Metallicities were derived from spectroscopy of HII regions as described in \cite{mapelli10}. Metallicity values for M81, NGC~1313, IC~342, NGC~2403 and M83 come from Mapelli et al. (2010, and references therein), while the metallicity of Ho~IX comes from \cite{makarova02}. For NGC~5204 and NGC~3034 no metallicity estimates are available that can be directly compared with those of the other galaxies. Column 10: Orbital period ($P$) as derived from observations. An estimate of the period is available only for NGC~3034 ULX5 and for NGC~1313 X-2. In the case of NGC~1313 X-2, $P$ might be either 6 d or 12 d  (\citealt{liu09}), depending on the relative importance of the variations induced by X-ray irradiation and ellipsoidal modulation of the donor. Column 11: $L_{\rm X,\,{}max}$ is the maximum observed X-ray luminosity. Column 12: references for $L_{\rm X,\,{}max}$: (1) \cite{winter06}, (2) \cite{pintore14}, (3) \cite{pintore12}, (4) \cite{feng10}.\\
In  columns 5--6: `NO' stands for `the source was not observed with this filter' (see G13 for details). The magnitudes provided in this table are Vega magnitudes and are corrected for Galactic extinction but not for intrinsic extinction.\label{tab:tab6}} 
%\end{flushleft}
%\end{center}
\end{deluxetable*}
%%%%%%%%%%%%%%%%%%%%%%%%%%%%%%%%%%%%%%%%%%%%%%%%%%%%%%%%%%%%%%%%%%%%%%%%%%%%%

 %%%%%%%%%%%%%%%%%%%%%%%%%%%%%%% TABLE 7%%%%%%%%%%%%%%%%%%%%%%%%%%%%%%%%%
\begin{deluxetable*}{lllllllll}
 \tabletypesize{\tiny}
 \tablewidth{0pt}
%\begin{center}
\tablecaption{Simulated systems that best match the selected ULX counterparts from G13} %\leavevmode
%\begin{scriptsize}
%\begin{tabular}[!h]{lllllllll}
%\hline
\tablehead{\colhead{Name} & \colhead{$\Delta{}_{\rm V}$} & \colhead{$\Delta{}_{\rm B-V}$}  & \colhead{$P$} & \colhead{$m_{\rm co}$}  & \colhead{$m_{\rm BH}$}  & \colhead{$Z$} & \colhead{Binary type} & \colhead{$L_{\rm Edd}$} \\
\colhead{} & \colhead{} &  \colhead{} & \colhead{(days)} & \colhead{(M$_\odot{}$)} & \colhead{(M$_\odot{}$)} & \colhead{(Z$_\odot{}$)} & \colhead{} & \colhead{$(10^{39}$ erg s$^{-1})$}} 
%\hline
\startdata
M81 X-6      & 0.05  & -0.05   & 8.13  & 152.5 &  16.0 & 0.01 & PB & 2.09  \\
M81 X-6      & -0.26 & 0.06    & 257   & 91.2  &  15.8 & 0.1  & PB & 2.05 \\
M81 X-6      & 0.03  & 0.06    &  80.8 & 48.2  &  7.9  & 1.0  & PB & 1.03 \\
M81 X-6      & 0.06  & -0.08   & 0.420 & 11.5  &  6.5  & 1.0  & PB & 0.84 \\
M81 X-6      & 0.31  & -0.04   & 8.51  & 58.4  &  6.0  & 0.01 & PB & 0.79 \vspace{0.1cm}\\
Ho~IX X-1     & -0.27 & 0.03    & 10.3  & 39.3  & 77.3  & 0.01 & EB & 10.05   \vspace{0.1cm}\\
%Ho~IX X-1     & -0.26 & 0.02    & 170   & 56.0  & 17.5  & 0.01 & PB & 2.28   \\
%Ho~IX X-1     & -0.50 & -0.06   & 61.1  & 47.9  &  13.1 & 0.01 & PB & 1.70  \\
%Ho~IX X-1     & 0.34  & 0.11    & 83.2  & 27.5  & 13.1  & 1.0  & PB & 1.70 \\
%Ho~IX X-1     & -0.16 & 0.03    & 334   & 42.4  & 11.6  & 0.1  & PB & 1.51   \\
%Ho~IX X-1     & -0.26 & -0.12   & 0.445 & 20.0  & 11.5  & 0.01 & PB & 1.50  \\
%Ho~IX X-1     & -0.26 & 0.13    & 402   & 49.4  & 10.5  & 1.0  & PB & 1.36  \\
%Ho~IX X-1     & -0.25 & -0.11   & 8.55  & 38.1  & 10.4  & 0.01 & PB & 1.35   \\
%Ho~IX X-1     & -0.49 & -0.17    & 0.792 & 7.2  & 9.6   & 1.0  & PB & 1.25  \\
%Ho~IX X-1     & 0.14  & 0.03    & 70.2  & 34.4  & 9.1   & 1.0  & PB & 1.18  \\
%Ho~IX X-1     &  0.33 & 0.01    & 155   & 53.6  & 8.5   & 0.1  & PB & 1.10  \\
%Ho~IX X-1     & -0.23 & -0.05   & 29.7  & 37.0  & 8.2   & 1.0  & PB & 1.07   \\
%Ho~IX X-1     & -0.30 & -0.06   & 26.6  & 35.7  & 8.1   & 1.0  & PB & 1.05  \vspace{0.2cm}\\
%Ho~IX X-1     & 0.20  & -0.11   & 12.2  & 46.3  & 7.4   & 0.1  & PB & 0.96  \\
%Ho~IX X-1     & 0.18  & -0.12   & 12.8  & 35.6  & 6.7   & 0.1  & PB & 0.88  \\
%Ho~IX X-1     & 0.32  & 0.12    & 408   & 38.3  & 6.5   & 1.0  & PB & 0.84  \\
%Ho~IX X-1     & -0.27 & -0.12   & 6.38  & 41.8  & 6.3   & 0.01 & PB & 0.82  \\
%Ho~IX X-1     & 0.40  & -0.02   & 112   & 31.0  & 6.1   & 0.1  & PB & 0.80 
%Ho~IX X-1     & 0.12   & -0.02  & 2024  & 30.6  & 5.2   & 0.1  & EB & 0.68 \vspace{0.2cm}\\
NGC~1313 X-1  & 0.06  & 0.15    & 6.98  & 36.2  & 12.3  & 1.0  & PB & 1.59 \\
NGC~1313 X-1  & -0.19 & 0.20    & 25.7  & 74.9  & 11.6  & 0.01 & PB & 1.51 \vspace{0.1cm} \\
%NGC~1313 X-1  & 0.36  & 0.17    & 11.1  & 59.0  & 11.3  & 0.01 & PB & 1.47 \vspace{0.1cm}\\
%NGC~1313 X-1  & 0.49  & 0.20    & 9.15  & 39.4  & 9.4   & 0.1  & PB & 1.22  \\
%NGC~1313 X-1  & 0.14  & 0.16    & 7.07  & 17.7  & 8.9   & 1.0  & PB & 1.15 \vspace{0.2cm}\\
NGC~1313 X-2  & 0.49  & 0.12    & 11.0  & 15.8  & 25.1  & 0.1  & EB & 3.26\\
NGC~1313 X-2  & -0.17 & 0.11    & 83.2  & 27.5  & 13.1  & 1.0  & PB & 1.70\\
NGC~1313 X-2  & 0.41  & -0.0009 &  25.7 & 74.9  & 11.6  & 0.01 & PB & 1.51 \\
NGC~1313 X-2  & 0.16  & 0.14    & 65.8  & 28.2  & 10.9  & 1.0  & PB & 1.41\\
NGC~1313 X-2  & 0.42  & 0.18    &  69.4 & 44.3  & 10.5  & 0.01 & PB & 1.36\\
NGC~1313 X-2  & 0.37  & 0.07    & 42.1  & 52.0  & 9.1   & 1.0  & PB & 1.19 \\
NGC~1313 X-2  & -0.37 & 0.03    & 70.2  & 34.4  & 9.1   & 1.0  & PB & 1.18\\
NGC~1313 X-2  & -0.18 & 0.01    & 155   & 53.6  & 8.5   & 0.1  & PB & 1.10\\
NGC~1313 X-2  & 0.37  & 0.04    & 46.8  & 25.7  & 7.6   & 0.1  & PB &  0.99\\
NGC~1313 X-2  & 0.38  & 0.18    & 375   & 24.7  & 7.5   & 0.01 & PB & 0.97 \\
NGC~1313 X-2  & -0.31 & -0.11   & 12.2  & 46.3  & 7.4   & 0.1  & PB & 0.96 \\
NGC~1313 X-2  & -0.33 & -0.12   &  12.8 & 35.6  & 6.7   & 0.1  & PB & 0.88 \\
NGC~1313 X-2  & -0.19 & 0.12    & 408   & 38.3  & 6.5   & 1.0  & PB & 0.84 \vspace{0.1cm}\\
%NGC~1313 X-2  & -0.11 & -0.02   & 112   & 31.0  & 6.1   & 0.1  & PB & 0.80 \vspace{0.1cm}\\
%NGC~1313 X-2  & 0.41  & 0.07    & 128   & 25.7  & 6.0   & 0.01 & PB & 0.78 &\\
%NGC~1313 X-2  & 0.09  & 0.11    & 291   & 29.4  & 5.3   & 1.0  & PB & 0.69 &\\
%NGC~1313 X-2  & -0.39 & -0.02   & 2024  & 30.6  & 5.2   & 0.1  & EB & 0.68 &\\
%NGC~1313 X-2  & 0.43  & 0.12    & 561   & 19.1  & 4.8   & 0.1  & PB & 0.62 &\\
%NGC~1313 X-2  & 0.02  & -0.09   & 6.00  & 26.7  & 4.7   & 1.0  & PB & 0.61 &\\
%NGC~1313 X-2  & 0.0006 & 0.03   & 113   & 19.7  & 4.7   & 1.0  & PB & 0.61 &\\
IC~342 X-1    & -0.27 & -0.15    & 1435  & 8.0   & 38.6  & 0.1  & EB & 5.02 \\
IC~342 X-1    & -0.30 & 0.03     & 1495  & 10.7  & 27.2  & 0.1  & EB & 3.54 \\
IC~342 X-1    & -0.19 & -0.03    & 1094  & 6.7   & 27.1  & 0.1  & EB & 3.53 \\
IC~342 X-1    & 0.04  & -0.19    & 194.5 & 8.1   & 24.6  & 0.1  & EB & 3.20 \\
IC~342 X-1    & 0.15  & -0.14    & 857   & 8.9   & 22.7  & 1.0  & EB & 2.96 \\
IC~342 X-1    & -0.19 & 0.08     & 3713  & 7.7   & 22.7  & 1.0  & EB & 2.95 \\
IC~342 X-1    & -0.10 & -0.15    & 140   & 6.7   & 22.5  & 0.1  & EB & 2.92 \\
IC~342 X-1    & 0.08  & -0.06    & 3007  & 9.7   & 22.3  & 0.1  & EB & 2.89 \\
IC~342 X-1    & -0.08 & 0.05     & 2976  & 9.5   & 21.3  & 1.0  & EB & 2.78 \\
IC~342 X-1    & -0.24 & 0.07     & 313   & 9.1   & 21.1  & 1.0  & EB & 2.74 \\
IC~342 X-1    & -0.25 & 0.12     & 1655  & 11.5  & 20.4  & 0.1  & EB & 2.66 \\
IC~342 X-1    & 0.49   & -0.04   & 6533  & 8.4   & 13.6  & 0.01 & EB & 1.77 \vspace{0.1cm}\\
NGC~2403 X-1  & -0.49 & 0.08    &  3916 & 15.1 & 6.3    &  0.01 & EB & 0.82 \\
NGC~2403 X-1  & -0.48 & 0.08    & 4221  & 15.0 &  6.2   &  0.01 & PB & 0.81\\
NGC~2403 X-1  & -0.24 & -0.06   &  60.5 & 7.2  & 6.0    &  0.01 & PB & 0.78 \\
NGC~2403 X-1  & -0.44 & 0.08    &  8113 & 9.1  & 5.9    &  0.01 & EB & 0.77 \\
NGC~2403 X-1  & -0.44 & 0.08    & 2768  & 16.3 & 5.9    & 1.0   & PB & 0.77 \\
NGC~2403 X-1  & -0.44 & 0.08    & 7616  & 13.9 & 5.9    & 1.0   & PB & 0.76 \\
NGC~2403 X-1  & -0.30 & -0.17   & 54.8  & 9.5  & 5.7    &  0.01 & PB & 0.75 \\
NGC~2403 X-1  & -0.34 & 0.08    &  6711 & 19.4 & 5.3    &  0.01 & PB & 0.69 \\
NGC~2403 X-1  & 0.03  & -0.19   & 35.4  & 9.9  & 5.1    & 0.01  & PB & 0.67 \\
NGC~2403 X-1  & -0.29 & 0.08    & 5227  & 12.5 & 5.0    & 1.0   & PB & 0.64 \\
NGC~2403 X-1  & -0.37 & 0.13    & 852   & 5.5  & 5.0    & 1.0   & EB & 0.64 \\
NGC~2403 X-1  & -0.24 & 0.07    & 354   & 8.1  & 4.8    & 0.1   & PB & 0.63 \vspace{0.1cm}\\
M83 XMM1     & -0.48 & -0.28   & 2007  & 18.5  & 6.2   & 1.0   & PB  & 0.80 \vspace{0.1cm}\\
NGC~5204 X-1  & 0.42  & -0.02   & 10.3  & 39.3 & 77.3   & 0.01  & EB & 10.05\\
NGC~5204 X-1  & 0.42  & -0.03   & 170   & 56.0 & 17.5   & 0.01  & PB & 2.28 \\
NGC~5204 X-1  & -0.06 & 0.08    & 138   & 99.0 & 16.4   & 0.1   & PB & 2.14 \\
NGC~5204 X-1  & -0.36 & -0.05   & 105   & 58.7 & 14.9   & 1.0   & PB & 1.94 \\
NGC~5204 X-1  & 0.19  & -0.10   & 61.1  & 47.9 & 13.1   & 0.01  & PB &  1.70 \\
NGC~5204 X-1  & -0.43 & 0.05    & 271   & 52.2 &  12.0  & 1.0   & PB & 1.56 \\
NGC~5204 X-1  & 0.43  & -0.17   & 0.445 & 20.0 & 11.5   & 0.01  & PB & 1.50 \\
NGC~5204 X-1  & -0.12 & -0.12   & 166   & 59.7 & 10.6   & 0.01  & PB & 1.37 \\
NGC~5204 X-1  & 0.43  & 0.09    & 402   & 49.4 & 10.5   & 1.0   & PB & 1.36 \\
NGC~5204 X-1  & 0.01  & -0.11   & 119   & 53.0 & 10.4   & 0.1   & PB & 1.35 \\
NGC~5204 X-1  & 0.44  & -0.16   & 8.55  & 38.1 & 10.4   & 0.01  & PB & 1.35  \\
NGC~5204 X-1  & -0.13 & -0.17   & 7.70  & 63.0 & 10.3   & 0.01  & PB & 1.34 \\
NGC~5204 X-1  & -0.22 & -0.17   & 9.33  & 60.3 & 10.1   & 0.01  & PB & 1.31 \\
NGC~5204 X-1  & -0.31 & -0.12   & 33.4  & 47.7 & 8.8    & 1.0   & PB & 1.14 \\
NGC~5204 X-1  & 0.46  & -0.10   & 29.7  & 37.0 & 8.2    & 1.0   & PB & 1.07 \\
NGC~5204 X-1  & 0.05  & -0.13   & 98.5  & 45.5 & 8.1    & 0.1   & PB & 1.06 \\
NGC~5204 X-1  & 0.38  & -0.11   & 26.6  & 35.7 & 8.1    & 1.0   & PB & 1.05 \\
NGC~5204 X-1  & -0.11 & -0.13   & 21.3  & 39.6 & 7.8    & 1.0   & PB & 1.02 \vspace{0.1cm}\\
%NGC~5204 X-1  & 0.41  & -0.17   & 6.38  & 41.8 & 6.3    & 0.01  & PB & 0.82 \\
%NGC~5204 X-1  & 0.13  & -0.16   & 6.51  & 43.0 & 5.7    & 0.01  & PB & 0.74 {0.2cm}\\
%%NGC~3034 ULX5 & 0.17  & -0.13   & 1636  & 14.9  & 17.0  & 0.1  & EB & 2.21 \\
NGC~3034 ULX5 & -0.03 & 0.05    & 5054  & 12.0  & 14.7  & 1.0  & EB & 1.91 
%NGC~3034 ULX5 & 0.50  & 0.10    & 3388  & 13.3  &  5.1  & 1.0  & PB & 0.67 \\
%\hline
%\noalign{\vspace{0.1cm}}
%\hline
%\end{tabular}
\enddata
%\end{scriptsize}
%\begin{flushleft}
%\footnotesize{
\tablecomments{{\it Notes.} Column 1: ULX name; column 2: $\Delta{}_{\rm V}\equiv{}V_{\rm sim}-V_{\rm obs}$, i.e. difference between simulated and observed $V$ magnitude; column 3: $\Delta{}_{\rm B-V}\equiv{}[(B_{\rm sim}-V_{\rm sim})-(B_{\rm obs}-V_{\rm obs})]$, i.e. difference between simulated and observed $B-V$ color; columns 4--9: period ($P$), mass of the donor star ($m_{\rm co}$), mass of the BH ($m_{\rm BH}$), metallicity ($Z$), binary type (i.e. PB or EB) and Eddington luminosity ($L_{\rm Edd}$) of the best-matching simulations, respectively.\\ %For NGC~1313 X-1 and M83 XMM1 we show only the best-matching  simulations. For the other six ULXs, 
We show all the best-matching simulations with $|\Delta{}_{\rm V}|\le{}0.50$, $|\Delta{}_{\rm B-V}|\le{}0.20$ and $L_{\rm Edd}\ge{}0.1\,{}L_{\rm X,\,{}max}$ (the adopted values of $L_{\rm X,\,{}max}$ are in Table~\ref{tab:tab6}). The best-matching simulations are listed in order of decreasing $L_{\rm Edd}$. See the text for details.\label{tab:tab7}}
%\end{flushleft}
%\end{center}
\end{deluxetable*}
%%%%%%%%%%%%%%%%%%%%%%%%%%%%%%%%%%%%%%%%%%%%%%%%%%%%%%%%%%%%%%%%%%%%%%%%%%%%%

From Figures~\ref{fig:fig6} and \ref{fig:fig7}, it is apparent that the observed counterparts populate the same regions in the CMD as the simulated ones. In Table~\ref{tab:tab7}, we list all the simulated systems that have Eddington luminosity $L_{\rm Edd}\ge{}0.1\,{}L_{\rm X,\,{}max}$ (where $L_{\rm X,\,{}max}$ is the maximum observed X-ray luminosity) and that differ from the observed counterparts by $|\Delta{}_{\rm V}|<0.5$ (where $\Delta{}_{\rm V}$ is the difference between simulated and observed $V$ magnitude) and by $|\Delta{}_{\rm B-V}|<0.2$ (where $\Delta{}_{\rm B-V}$ is the difference between simulated and observed $B-V$ color). The requirement that $L_{\rm Edd}\ge{}0.1\,{}L_{\rm X,\,{}max}$ is necessary because  our models have been calculated under the assumption of sub-Eddington accretion: a strongly super-Eddington accretion and/or the presence of beaming would affect the optical counterpart significantly and cannot be accounted for by our current models. The tolerance ranges in $V$ and $B-V$ account for the fact that some observed ULX counterparts are known to vary significantly (e.g. NGC~1313 X-2, \citealt{mucciarelli07}; \citealt{grise08}; \citealt{liu09}; Ho IX X-1, \citealt{grise11}).

We do not find any simulated systems that satisfy the aforementioned requirements in the case of both M83 XMM1 and NGC~3034 ULX5. M83 XMM1 is one of the three sources for which the $V$ magnitude was derived from the F606W filter, likely introducing a larger uncertainty.  In Table~\ref{tab:tab7}, we report the simulated system which is closer to the observed counterpart of M83 XMM1: it has $\Delta{}_{\rm B-V}=-0.28$ and $\Delta{}_{\rm V}=-0.48$. 

In the case of NGC~3034 ULX5, all the simulated systems that match the criteria on optical magnitude and color have $L_{\rm Edd}\ll{}0.1\,{}L_{\rm X,\,{}max}$. In Table~\ref{tab:tab7}, we report the simulated system which is closer to the observed counterpart according to $\Delta{}_{\rm B-V}$ and $\Delta{}_{\rm V}$, but has $L_{\rm Edd}\sim{}0.02\,{}L_{\rm X,\,{}max}$. NGC~3034 ULX5 (also known as M82 X-1) is an outlier of our sample, under many respects. First, the intrinsic extinction is expected to be very high, but cannot be reliably quantified (G13). Second, the best-matching simulated system has much longer orbital period (5000 d)  than the observed one (62 d, \citealt{kaaret06}). Third, the maximum observed X-ray luminosity is by far the highest in the sample (\citealt{feng10}). This source likely belongs to a completely different class (an intermediate-mass BH has been required to explain some of its properties, e.g. \citealt{kaaret01, patruno06, casella08, feng10}).

In the other cases, we come to some remarkable results. The counterparts of M81 X-6 and NGC~1313 X-1 are matched only by simulated RLO-PBs, the counterparts of NGC~1313 X-2, NGC~2403 X-1 and  NGC~5204 X-1 are matched by both simulated  RLO-PBs and simulated RLO-EBs, while the counterparts of Ho~IX X-1 and IC~342 X-1 are matched only by simulated  RLO-EBs.  Considering the three observed counterparts that are matched by both  RLO-PBs and  RLO-EBs, the RLO-EBs always have the highest $L_{\rm Edd}$ (because they host more massive BHs). Thus, the  RLO-EBs have the advantage that they do not require an exceedingly high super-Eddington factor in these three cases.

If we require not only a good matching between observed and simulated optical counterparts, but also that $L_{\rm Edd}\ge{}L_{\rm X,\,{}max}/3$ (i.e. only a mild super-Eddington factor\footnote{We chose to focus on mildly super-Eddington systems not because we think that the considered ULXs cannot be strongly super-Eddington, but just because our code assumes Eddington-limited accretion. %Using our code to study super-Eddington systems is not a reasonable approximation. 
In a forthcoming study, we will generalize our results to super-Eddington systems.}), %the maximum observed X-ray luminosity does not exceed the Eddington luminosity by a factor larger than three, 
then Ho~IX X-1, NGC~1313 X-2, IC~342 X-1 and (marginally) NGC~5204 X-1 can be matched only by simulated RLO-MSBHs. The mass of the best-matching MSBHs is in the $25-77$ M$_\odot{}$ range. In the following, we focus on these four sources.

%The observed counterparts of Ho~IX X-1, NGC~1313 X-2, IC~342 X-1 and  NGC~5204 X-1 can be matched by simulated RLO-MSBH systems, with MSBH mass in the $25-77$ M$_\odot{}$ range. 

\subsubsection*{Ho~IX X-1}
In the case of Ho~IX X-1, a MSBH accretor is particularly favored by the high observed X-ray luminosity ($L_{\rm X,\,{}max}=2.8\times{}10^{40}$ erg s$^{-1}$). The best-matching RLO binary\footnote{The second best-matching system for Ho~IX X-1  is a RLO-PB at $Z=0.01$ Z$_\odot$, hosts a BH with $m_{\rm BH}=17.5$ M$_\odot$, a donor star with  $m_{\rm co}=56.0$ M$_\odot$ and has orbital period $P=170$ d. It  is not shown in Table~\ref{tab:tab7}, because it has $L_{\rm Edd}<0.1\,{}L_{\rm X,\,{}max}$.} hosts a MSBH with $m_{\rm BH}=77.3$ M$_\odot$, but has a metallicity much lower than the observed metallicity of Ho~IX ($Z\sim{}0.4$ Z$_\odot$, Table~\ref{tab:tab6}).

\subsubsection*{NGC~1313 X-2}
The counterpart of NGC~1313 X-2 is particularly well matched by a simulated RLO-MSBH not only for the high observed X-ray luminosity ($L_{\rm X,\,{}max}=0.8\times{}10^{40}$ erg s$^{-1}$), but also for the orbital period. In fact, the orbital period of the simulated RLO-MSBH system with $m_{\rm BH}=25.1$ M$_\odot{}$ is $P=11$ d (Table~\ref{tab:tab7}). NGC~1313 X-2 has an observed period either $P=6$ d or $P=12$ d (\citealt{liu09}), depending on the relative importance of the variations induced by X-ray irradiation and ellipsoidal modulation of the donor (Table~\ref{tab:tab6}). 
%depending on whether the  X-ray irradiation of the donor is important or negligible. 
The simulated RLO-MSBH system with $P=11$ d well matches the latter case. This is consistent with what reported by \cite{patruno10}, who found that, for an orbital period of $\sim{}12$ days, a $\sim{}20$ M$_\odot$ (or slightly larger) BH with a $\sim{}12-15$ M$_\odot$ H-shell burning donor is compatible with the observed photometry in case of isotropic X-ray irradiation. Furthermore, the metallicity of this simulated RLO-MSBH ($Z=0.1$ Z$_\odot$) is close to the metallicity in proximity of NGC~1313 X-2 ($Z\sim{}0.1$ Z$_\odot$, \citealt{ripamonti11} and Table~\ref{tab:tab6}).

\subsubsection*{IC~342 X-1}
IC~342 X-1 is the observed source most significantly matched by a RLO-MSBH binary. IC~342 X-1 has an X-ray luminosity of $\sim{}10^{40}$ erg s$^{-1}$, is close to a star-forming region, is associated with an ionized nebula and has two possible counterparts (\citealt{feng08}; \citealt{cseh12}). The brightest one is consistent with a F8 to G0 Ib supergiant ($\gtrsim{}10$ Myr old), if no disk emission is considered. In contrast, if the disk contribution is important, almost no constraints can be put on the companion (\citealt{feng08}).

In our simulations, the best-matching models of IC~342 X-1 have $m_{\rm BH}=21-39$ M$_\odot$, corresponding to an Eddington luminosity $L_{\rm Edd}=3-5\times{}10^{39}$ erg s$^{-1}$. This requires a mild $3-4$ super-Eddington factor to match the observed $L_{\rm X,\,{}max}$. The companion star has a mass $m_{\rm co}=6.7-10.7$ M$_\odot$ and is a red super-giant star (with $T_{\rm eff}\sim{}3500$ K and $L\sim{}1-6\times{}10^4$ L$_\odot$). The RLO phase starts at $t=25-60$ Myr, depending on the simulation. 
%Five out of six best-matching simulated systems have $F_{\rm disk}=1.0-1.8\,{}F_{\rm co}$, while the remaining system has $F_{\rm disk}=52\,{}F_{\rm co}$. Thus, the contribution of the irradiated disk is important. 

Most of the best-matching models are at $Z=0.1$ Z$_\odot{}$, which is fairly close to the observed metallicity of IC~342 ($Z\sim{}0.1-0.4$ Z$_\odot{}$, Table~\ref{tab:tab6}). Finally, in comparing the observations with our best-matching model, we do not account for intrinsic extinction. This is consistent with the observations, since \citet{feng08} point out that the local absorption is not dominant. %Galactic extinction is the dominant component.

\subsubsection*{NGC~5204 X-1}
The best-matching RLO-MSBH system for NGC~5204 X-1 has $L_{\rm Edd}>L_{\rm X,\,{}max}$, while there are two RLO-LBH systems that require just a mild super-Eddington factor of $\sim{}4$.

\subsection{Caveats and future work}\label{sec:caveat}

 In this section, we discuss the possible issues concerning our models and the comparison with the observed ULX counterparts. First, we recall that our results hold only if the observed ULXs were born in a dense stellar association/young SC. The simulations presented in this paper do not predict the evolution of BHs that were born and evolved in low-density regions. On the other hand, most stars (and especially the most-massive stars) are believed to form in young SCs, supporting the assumption that  most BHs form in young SCs. 
%and dynamically evolve 
 We notice that at least four sources whose counterparts are best-matched by RLO-EBs are associated with star forming regions: Ho~IX X-1 and NGC~1313 X-2 are in relatively loose young SCs (\citealt{ramsey06}; \citealt{grise08}), IC~342 X-1 is in a star forming region (\citealt{feng08}) and NGC~3034 ULX5 is %displaced by 
0.65 arcsec away from a massive young SC (\citealt{voss11}).
%RIC~ORDARSI DI AGGIUNGERE LA CONNESISONE CON STAR CLUSTERS (VOSS)
%%%%%%%%%%%%%%%%%%%%%%%%%%%%%%%%%%% FIGURE 9 %%%%%%%%%%%%%%%%%%%%%%%%%%%%%%%%%%
\begin{figure}
\center{{
\epsfig{figure=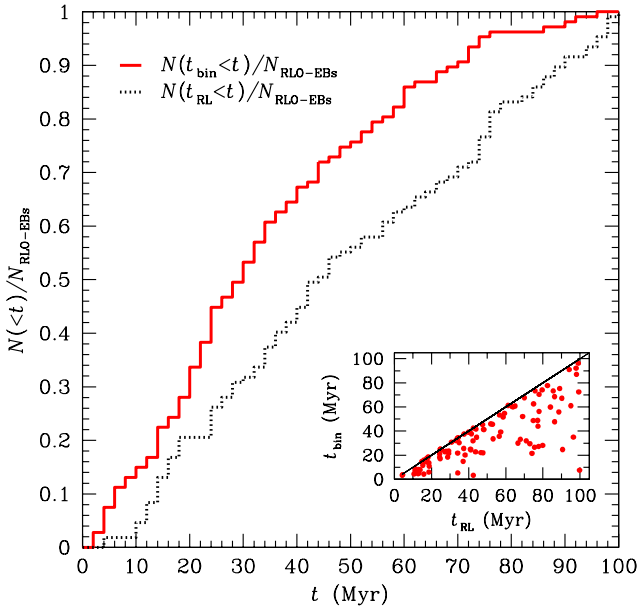,width=8.5cm} %was fig9.eps 
\caption{\label{fig:fig9}
%Cumulative distributions of simulated RLO-EBs. 
Red solid line ($N(t_{\rm bin}<t)/N_{\rm RLO-EBs}$): cumulative distribution of the formation times ($t_{\rm bin}$) of the EBs that will undergo RLO during the simulation, normalized to the total number of RLO-EBs ($N_{\rm RLO-EBs}$). Black dotted line ($N(t_{\rm RL}<t)/N_{\rm RLO-EBs}$): cumulative distribution of the times when the first RLO phase starts ($t_{\rm RL}$), for the simulated EBs that undergo RLO, normalized to the total number of RLO-EBs ($N_{\rm RLO-EBs}$). In the inset: $t_{\rm bin}$ versus $t_{\rm RL}$ for the simulated RLO-EBs. The solid black line marks the points with  $t_{\rm bin}=t_{\rm RL}$.}}}
\end{figure}
%%%%%%%%%%%%%%%%%%%%%%%%%%%%%%%%%%%%%%%%%%%%%%%%%%%%%%%%%%%%%%%%%%%%%%%%%%%%%%

A major issue is represented by infant mortality and tidal disruption of SCs. The fraction of SCs that survive gas evaporation is highly uncertain ($\approx{}5-30$ per cent, e.g. \citealt{gieles11}). Disruption of SCs from the tidal field of the host galaxy is another important ingredient. Overall, the fraction of young SCs that survive for more than $\sim{}100$ Myr is uncertain, but it is probably very low ($\sim{}5$ per cent, \citealt{lada03}). The simulations presented in this paper account neither for gas evaporation nor for the galactic tidal field, which will be considered in forthcoming studies. On the other hand, we can indirectly estimate the impact of SC mortality on our results by looking at the time when EBs form ($t_{\rm bin}$) and start the RLO phase ($t_{\rm RL}$). %$t_{\rm bin}$ is defined as the time when the last dynamical exchange occurs before the binary enters a RLO phase.

The underlying idea is that an EB cannot exist, if the host SC is disrupted or evaporates before the formation of the EB by dynamical exchange (i.e. before $t_{\rm bin}$). After the formation of the EB, flyby encounters with other stars can occur and they may contribute to trigger the RLO phase. On the other hand, flybys are generally much less important than exchanges, in the sense that stellar evolution is the main driver of the RLO phase in most EBs after the initial exchange. Thus, we can optimistically (pessimistically) assume that an EB enters the RLO phase if the SC survives for a time $t>t_{\rm bin}$  ($t>t_{\rm RL}$). Fig.~\ref{fig:fig9} shows the cumulative distribution of $t_{\rm bin}$ and $t_{\rm RL}$ for all the simulated RLO-EBs, without distinguishing for metallicity. From this Figure, it is apparent that $\sim{}50$ per cent ($\sim{}30$ per cent) of EBs that will undergo RLO have already formed (have already gone through the first RLO phase) at $t\sim{}30$ Myr. 
%At $t\sim{}6$ Myr, i.e. after the collapse of the SC core, $\sim{}10$ per cent ($\sim{}2$ per cent) of the EBs that will undergo RLO have already formed (already started RLO). 
These percentages indicate that the number of RLO-EBs is significantly quenched when the SC is disrupted during an early evolutionary stage. Thus, our results represent a robust upper limit to the statistics of RLO-EBs in intermediate-mass young SCs.

%conclude that a RLO-EB exists if the SC is disrupted after $t_{\rm RL}$

%If the EB can form, it is still possible (although not frequent) that the EB cannot evolve into RLO-EB

A further caveat is that our results are based on different realizations of a unique SC model, with the same PB fraction ($f_{\rm PB}$), total SC mass ($M_{\rm TOT}$), initial core density and concentration (see Table~\ref{tab:tab1}). 

$f_{\rm PB}$ is a delicate ingredient of our simulations. Recent observations show that the observed binary fraction can be very high in open and young SCs ($\gtrsim{}30\,{}$ per cent, with a large uncertainty, e.g. \citealt{sollima10}; \citealt{li13}). On the other hand, PBs are a bottleneck for direct-summation N-body codes. Thus, most direct-summation N-body simulations do not include PBs (or include a low fraction of PBs). In our simulations, we assume $f_{\rm PB}=0.1$, which implies (according to our definition of $f_{\rm PB}$, see Section~\ref{sec:sec2.1}) that 18 per cent of stars are initially in binaries. 

In Appendix~\ref{appendix:fPB}, we discuss the results of test simulations  with $f_{\rm PB}=0.2$ (corresponding to 33 per cent of stars in primordial binaries). By comparing   the case with $f_{\rm PB}=0.1$ and  the case with $f_{\rm PB}=0.2$, we find that the total number of RLO-EBs remains unchanged for different values of $f_{\rm PB}$, while the total number of RLO-PBs grows about linearly with $f_{\rm PB}$. Thus, we conclude that the total number of RLO-EBs found in our simulations does not significantly depend on $f_{\rm PB}$, while the fraction of RLO-EBs with respect to the total number of RLO binaries does. %This implies that the number of RLO-EBs depends on the structural SC properties (e.g. mass, central density and virial)

%In a forthcoming study, we will focus on the impact of PBs on the formation of X-ray binaries and other exotic binaries. %In this paper, we briefly discuss the effects of a higher PB fraction in Appendix~\ref{appendix:fPB}. The main result of 

In a forthcoming study, we will focus on the impact of the main structural parameters of young SCs (e.g. total SC mass, initial core density and concentration) on the formation of X-ray binaries and other exotic binaries. We expect that the initial core density is a crucial parameter for the formation of EBs and thus for RLO-EBs, because a higher density implies a higher three-body encounter rate (e.g. \citealt{sigurdsson93}). The role of the total SC mass is more difficult to predict, because a larger mass means a larger number of binaries per single SC but also a longer two-body relaxation timescale. Furthermore, SCs with larger mass are less numerous in the local Universe, but might avoid infant mortality and tidal disruption (depending on the potential well of the host galaxy).

%In a forthcoming study, we will focus on the impact of  on the formation of X-ray binaries and other exotic binaries.
%cluster mass core density W0

Finally, the comparison with observations presented here is still preliminary. The main limitations are that the accretion disk is assumed to be a Shakura-Sunyaev disk and the accretion rate is Eddington limited. Recent studies (e.g. \citealt{gladstone09}; \citealt{sutton13}; \citealt{middleton14}; \citealt{pintore14}) indicate that super-Eddington accretion might explain a large fraction of ULXs. A more detailed photometric analysis and modeling of disk emission at super-Eddington rates (involving a major upgrade of the \citealt{patruno10} code), combined with a new set of N-body simulations, is under way.

% (for a standard accretion efficiency of $\sim{}0.1$), the luminosity is limited at Eddington and the excess mass is assumed to be expelled from the system.(other disk models have not been implemented yet) and 

\section{Conclusions}
We simulated the evolution of 600 young SCs with different metallicity ($Z=0.01$, 0.1 and 1 Z$_\odot$), to investigate the demographics of RLO systems powered by BHs in young SCs. The properties of the simulated young SCs match those of intermediate-mass dense young SCs in the Milky Way. Thus, the simulated young SCs are dynamically active: they undergo core collapse in $\sim{}3$ Myr, and their evolution is driven by three-body encounters and dynamical exchanges.

In this paper, we focus on the 244 simulated BH binaries that undergo RLO (since each simulation lasts for 100 Myr, this implies a formation rate of $\sim{}0.004$ Myr$^{-1}$ RLO-BH binaries per SC). About 44 and 56 per cent of the simulated RLO BH binaries are RLO-EBs and RLO-PBs, respectively (Table~\ref{tab:tab4}). The properties of RLO-EBs are markedly different from those of RLO-PBs. RLO-EBs are generally powered by more massive BHs and start the RLO phase later than RLO-PBs (see Fig.~\ref{fig:fig1}). As a consequence, the mass of the donor star in RLO-EBs is $\le{}20$ M$_\odot{}$, generally smaller than the mass of the  donor star in RLO-PBs (Fig.~\ref{fig:fig2}). Most donor stars ($\sim{}43$ per cent) are MS stars, mainly close to the TAMS. Red giant and red super-giant branch stars are also common ($\sim{}33$ per cent, Table~\ref{tab:tab5}). 

RLO binaries powered by MSBHs ($\ge{}25$ M$_\odot$) are almost a sub-class of RLO-EBs, since all RLO-MSBHs but one  form through dynamical exchange (Figures~\ref{fig:fig3} and \ref{fig:fig4}). This confirms that it is very difficult for PBs to evolve into RLO-MSBHs (see e.g. \citealt{linden10}), whereas dynamical exchanges are very efficient in producing RLO-MSBHs (see M13). 
%the idea that MSBHs born in PBs can hardly power RLO systems (see e.g. \citealt{linden10}), whereas dynamical exchanges are very efficient in producing RLO-MSBH systems (see M13). 

MSBHs form mainly from the direct collapse of massive metal-poor stars, but nine MSBHs in RLO systems (corresponding to $\sim{}21$ per cent of all the RLO-MSBHs) formed from the merger of either two stars or a BH and a star.

In this paper, we produced CMDs of the counterparts of the simulated X-ray binaries. To this purpose, we used the code by Patruno \&{} Zampieri (2008, 2010), which couples the emission from the star with the contribution of the accretion disk. RLO-EBs and RLO-PBs form two different populations in the CMD. The sequence defined by  $V\approx{}24.3\,{}-\,{}1.6\,{}(B-V)$ is mainly populated by RLO-EBs (Fig.~\ref{fig:fig6}), whose donor stars are predominantly older and less-massive than those of RLO-PBs. In contrast, RLO-PBs are mostly associated with the bluer counterparts. In both cases, there are some remarkable exceptions (i.e. very blue RLO-EBs and very red RLO-PBs).  RLO-MSBHs  populate the same sequence as RLO-EBs. On the other hand,  RLO-MSBHs represent the high-luminosity tail of  RLO-EBs, since they populate mainly the upper envelope of this sequence (Fig.~\ref{fig:fig7}).

%donors (RLO-EBs vs RLO-PBs)

%MSBHs

%donor

These results provide important insights to understand the observations of bright X-ray binaries in star forming regions, such as the ULXs. ULXs are associated with star forming regions and/or young SCs, and seem to anti-correlate with the metallicity of the host galaxy (e.g. \citealt{mapelli10}; \citealt{mapelli11a}). We compared the simulated RLO systems with nine of the ULX counterparts listed in G13 (Table~\ref{tab:tab6}). The observed counterparts populate the same regions of the CMD as the simulated ones (Fig.~\ref{fig:fig6}). %Five out of nine observed ULX counterparts (M81 X-6, Ho~IX X-1, NGC~1313 X-1, NGC~1313 X-2, NGC~5204 X-1) have blue colors ($-0.4\le{}B-V\le{}-0.1$), while the remaining four systems (IC~342 X-1, NGC~2403 X-1, M83 XMM1, NGC~3034 ULX5) match the $V\approx{}24.3\,{}-\,{}1.6\,{}(B-V)$ sequence (Fig.~\ref{fig:fig6}). 

 In particular, the counterparts of M81 X-6 and NGC~1313 X-1 are matched only by simulated RLO-PBs, the counterparts of NGC~1313 X-2, NGC~2403 X-1 and  NGC~5204 X-1 can  be matched by both simulated  RLO-PBs and simulated RLO-EBs, while the counterparts of  Ho~IX X-1  and IC~342 X-1  are matched only by simulated  RLO-EBs. %The counterparts of M83 XMM1 and NGC~3034 ULX5 cannot be satisfactorily matched by any symulated system.

If we require not only a good matching between observed and simulated optical counterparts, but also that the maximum observed X-ray luminosity does not exceed the Eddington luminosity by a factor larger than three, then Ho~IX X-1, NGC~1313 X-2, IC~342 X-1 and (marginally) NGC~5204 X-1 can be matched only by simulated RLO-MSBHs.

The counterpart of NGC~1313 X-2 is particularly well matched by a simulated RLO-MSBH not only for the high observed X-ray luminosity ($L_{\rm X,\,{}max}=0.8\times{}10^{40}$ erg s$^{-1}$), but also for the %fact that a simulated RLO-MSBH system (with $m_{\rm BH}=25.1$ M$_\odot{}$) has 
simulated orbital period ($P=11$ d, see Table~\ref{tab:tab7}), consistent with the observed one in case ellipsoidal modulations significantly affect the light curve (\citealt{liu09}).
%very close to the observed one (in case of negligible irradiation, \citealt{liu09}). }
%the counterparts of M81 X-6, NGC~1313 X-2, NGC~2403 X-1 and  NGC~5204 X-1 are best-matched by simulated  RLO-PBs, while the counterparts of IC~342 X-1 and NGC~3034 ULX5 (i.e. the two redder systems) are matched by simulated  RLO-EBs.  The counterpart of  Ho~IX X-1, which is rather blue, can be matched by both  RLO-PBs and  RLO-EBs. In most cases, the mass of the BH in the best-matching simulated system is $<20$ M$_{\odot{}}$, requiring super-Eddington accretion to match the X-ray luminosity.

The counterpart of IC~342 X-1 is the most significantly matched by a RLO-MSBH (with BH mass $m_{\rm BH}=21-39$ M$_{\odot{}}$).  This interpretation is supported also by  recent {\it Chandra}, NuSTAR and XMM observations (\citealt{marlowe14}; \citealt{rana14}; \citealt{pintore14}). The donor star of the best matching systems is a relatively low-mass (6.7--10.7 M$_\odot$) red super-giant star.

 Unfortunately, a robust identification of the optical counterpart was obtained only for a few ULXs (e.g. G13 and references therein). Furthermore, a measurement of the orbital period is available only for four ULXs (NGC~3034 ULX5, \citealt{kaaret06}; NGC~1313 X-2, \citealt{liu09}; CXOU J123030.3+413853  in NGC~4490, \citealt{esposito13}; and M101 X-1, \citealt{liu13}). Besides, in our study we did not consider intrinsic extinction, which is an additional factor of uncertainty.  In addition, our code (\citealt{patruno10}) is based on a Shakura-Sunyaev disk model, and assumes that any mass transfer in excess of the Eddington limit is expelled from the system. Thus, it is very difficult to make  a robust comparison between data and models.

In addition, ULXs are very peculiar and rare objects (an occurrence rate of less than one ULX per galaxy has been found in the catalog by \citealt{swartz11}). Our simulations represent a statistically limited sample: 600 young SCs are approximately the young SC content of a single starburst galaxy.  Furthermore, we considered only a relatively small fraction of PBs, we simulated different realizations of a single SC model (with the same total mass, virial radius and concentration), and we did not account for gas evaporation and for the tidal field of the host galaxy.  Thus, a larger sample of $N-$body simulations (including prescriptions for the missing ingredients) is necessary to obtain a complete statistical description of RLO systems in young SCs and to compare them with observed X-ray sources.

Bearing these caveats in mind, our simulations show (for the first time in a self-consistent way) that RLO-MSBHs in young SCs can reproduce the optical properties of observed ULX counterparts. 
%The idea that some ULXs are powered by MSBHs was originally proposed by Mapelli et al. (2009, see also \citealt{zampieri09}; M13).  
 This result is very promising, because a growing number of observations suggest that (some) ULXs might be powered by MSBHs: this scenario is consistent with   the observed anti-correlation between ULXs and metallicity of the host galaxy (\citealt{mapelli09}; \citealt{mapelli10}; \citealt{mapelli11a}; \citealt{prestwich13}), with the measured mass function of M~101 ULX-1 (\citealt{liu13}), and with the recent radio observations of Holmberg~II X-1 (\citealt{cseh14}).

\section*{Acknowledgments}
 We thank the anonymous referee for their comments, which significantly improved the paper. We also thank Jeannette Gladstone and Nathan Leigh for their invaluable comments, and Alessia Gualandris, Alessandro Bressan and Emanuele Ripamonti for useful discussions. We made use of the {\sc starlab} (version 4.4.4) public software environment and of the SAPPORO library (\citealt{gaburov09}). We thank the developers of {\sc starlab}, and especially its primary authors (P.~Hut, S.~McMillan, J.~Makino, and S.~P.~Zwart). We thank the authors of SAPPORO, and in particular E. Gaburov, S. Harfst and S.  Portegies Zwart. %We also thank Paola Marigo, Alessia Gualandris, Antonella Vallenari, Rosanna Sordo, Laura Greggio, Mauro Barbieri and Monica Colpi for stimulating discussions.
 We acknowledge the CINECA Award N.  HP10CXB7O8, HP10C894X7, HP10CGUBV0, HP10CP6XSO and HP10C3ANJY for the availability of high performance computing resources and support.  The simulations were run on the graphics processing unit (GPU) clusters IBM-PLX and EURORA at CINECA. %The processors available on PLX are six-cores Intel Westmere 2.40 GHz (two per node), while the GPUs are NVIDIA Tesla M2070 and M2070Q (two per node). Each single job ran over one processor and two GPUs and required 50 CPU hours on average.
MM and LZ acknowledge financial support from the Italian Ministry of Education, University and Research (MIUR) through grant FIRB 2012 RBFR12PM1F, and from INAF through grant PRIN-2011-1. MM acknowledges financial support from CONACyT through grant 169554.
% MM thanks the Aspen Center for Physics, where part of this work was done.

\begin{appendix}
\section{The sample of ULX counterparts}\label{appendix:counterparts}
 The most recent and homogeneous sample of optical ULX counterparts has been reported by G13. Of the 33 ULXs observed with both {\it HST} and {\it Chandra}, and studied in G13, nine have no visible counterparts, and two were found to lie too close to the nucleus of the host galaxy to be classified as ULXs. The remaining 22 ULXs have at least one detected possible counterpart with {\it HST}. Among them, we selected nine ULXs (see Table~\ref{tab:tab6}), according to the following criteria. 

We selected the six ULXs (M81 X-6, Ho~IX X-1, NGC~1313 X-1, NGC~1313 X-2, IC~342 X-1 and NGC~3034 ULX5) with {\it HST} data in the F555W and in the F435W filters (approximately corresponding to the Johnson $V$ and $B$ filters, which are implemented in the code by \citealt{patruno10}). 
 In addition, we selected the three ULXs (NGC~2403 X-1, M83 XMM1 and NGC~5204 X-1) for which there are {\it HST} data in both the F435W and the F606W filter (corresponding to a wide $V$ filter), but no observations with the  F555W filter. Three of the six ULXs with  {\it HST} data in both the F555W and the F435W filters have also observations in the  F606W filter (M81 X-6, NGC~1313 X-1 and  IC~342 X-1). 
 As a proxy to the transformation between different filters, for the three sources that lack F555W data we convert the F606W magnitude into a F555W magnitude using the average shift between the F555W and F606W filters, determined from the counterparts with both measurements.

 Five of the nine considered sources (Ho~IX X-1, NGC~1313 X-2, IC~342 X-1, NGC~5204 X-1 and NGC~3034 ULX5) have multiple possible counterparts in the {\it Chandra} error box. In particular, both IC~342 X-1 and NGC~5204 X-1 have two possible counterparts in the {\it Chandra} error box and each of them has {\it HST} data in both $V$ and $B$. In Figures~\ref{fig:fig6} and \ref{fig:fig7}, we plot  only the counterpart of IC~342 X-1 and that of NGC~5204 X-1 that were labelled as 1 in Table~4 of G13. % and column 2 of our Table~4). 
We analyzed also the counterparts of IC~342 X-1 and of NGC~5204 X-1 that were labelled as 2, but they do not match any simulated systems.
  
 Both NGC~1313 X-2 and NGC~3034 ULX5 have two possible counterparts in the {\it Chandra} error box, but only one of the two was detected in both $V$ and $B$. Thus, for both NGC~1313 X-2 and NGC~3034 ULX5, we consider only the counterpart which has been detected in both $V$ and $B$. Furthermore, object 1 is the most likely counterpart of NGC~1313 X-2, based on the identification of the He~II~4686~\AA{} emission line in the spectrum  (\citealt{grise08}) and on the accurate {\it Chandra} and {\it HST} astrometry (\citealt{liu07}). In addition, object 2 might be an artifact of data analysis (Jeannette Gladstone, private communication).

 Ho~IX X-1 has three possible counterparts in the {\it Chandra} error box: counterparts 1 and 3  have F435W and F555W data, while counterpart 2 was detected only in the F330W filter (because the data considered in G13 do not have the spatial resolution to separate it from source 1 in any band but F330W).  In Figures~\ref{fig:fig6} and \ref{fig:fig7}, we plot  only counterpart 1 of  Ho~IX X-1. We analyzed also counterpart 3, but it does not match any simulated system. Furthermore, object 1 is the most likely counterpart of Ho~IX X-1, based on the identification of the He~II~4686~\AA{} emission line in the spectrum  (\citealt{roberts11}).

%For these reasons, in Table~4, we report only the counterparts that are shown in Figures~\ref{fig:fig6} and \ref{fig:fig7} and that are considered in the following analysis.

\section{The impact of the binary fraction}\label{appendix:fPB}
 In this Appendix, we discuss the impact of the PB fraction on our results.  In the runs presented in the main text, we have assumed $f_{\rm PB}=0.1$, which means that 18 per cent of the simulated stars are in binaries at the beginning of the simulation. Recent observations (e.g. \citealt{sollima10}; \citealt{li13}) indicate that the binary fraction in young SCs may be significantly higher. However, high PB fractions are a severe bottleneck for  direct-summation N-body simulations. Thus, it is very difficult to obtain a good statistical sample of simulated young SCs with a high PB fraction. 

In order to quantify the impact of the binary fraction on our results, we have run an additional sample of 100 young SCs with metallicity $Z=0.1$ Z$_\odot$ and PB fraction $f_{\rm PB}=0.2$. This means that $\sim{}33$ per cent of the stars are members of a binary system at the beginning of the simulation. 

%It is difficult to compare SCs with different binary fraction, because the relevant dynamical timescales (e.g. dynamical-friction timescale and core-collapse timescale) may be different. On the other hand, 
%In comparing SCs with different binary fraction one should account for the fact that the relevant dynamical timescales (e.g. dynamical-friction timescale and core-collapse timescale) may be different. For this reason, 

To compare SCs with the same relevant dynamical timescales (e.g. initial two-body relaxation timescales), we imposed that the SCs with $f_{\rm PB}=0.2$ have the same total mass ($M_{\rm TOT}\sim{}3500$ M$_\odot{}$) and virial radius (1 pc) as the SCs with $f_{\rm PB}=0.1$. %Thus, the initial two-body relaxation timescales are very similar. 
More details about the initial conditions of the SCs with $f_{\rm PB}=0.2$ are given in Table~\ref{tab:tab8}.

%%%%%%%%%%%%%%%%%%%%%%%%%%%%%%% TABLE A1%%%%%%%%%%%%%%%%%%%%%%%%%%%%%%%%%
\begin{deluxetable}{ll}
\tabletypesize{\tiny}
 \tablewidth{0pt}
\tablecaption{Most relevant initial conditions of 100 runs with $f_{\rm PB}=0.2$.} %\leavevmode
\tablehead{
\colhead{Parameter} 
& \colhead{Values}}
\startdata
$W_0$ & 5 \\
$N_\ast{}$ & 5640 \\
$r_{\rm c}$ (pc) & 0.4\\
$c$ & 1.03\\
IMF & Kroupa (2001)\\
$m_{\rm min}$ (M$_\odot{}$) & 0.1\\
$m_{\rm max}$ (M$_\odot{}$) & 150\\
$Z\,{}({\rm Z}_\odot{})$ & 0.1\\
$f_{\rm PB}$ & 0.2
%\noalign{\vspace{0.1cm}}
%\hline
%\end{tabular}
\enddata
%\begin{flushleft}
\tablecomments{
%\footnotesize{
{\it Notes.} The symbols are the same as in Table~\ref{tab:tab1}. In each simulated SC, there are initially 4700 CMs, among which 940 are designated as `binaries' and 3760 are `single stars'. Thus, 1880 stars per SC are initially in binaries.\label{tab:tab8}}
%\end{flushleft}
%\end{center}
%\end{table}
\end{deluxetable}
%%%%%%%%%%%%%%%%%%%%%%%%%%%%%%%%%%%%%%%%%%%%%%%%%%%%%%%%%%%%%%%%%%%%%%%%%%%%%
%%%%%%%%%%%%%%%%%%%%%%%%%%%%%%% TABLE A2%%%%%%%%%%%%%%%%%%%%%%%%%%%%%%%%%
\begin{deluxetable}{llllll}
\tabletypesize{\tiny}
 \tablewidth{0pt}
\tablecaption{Comparison of the simulated RLO systems in SCs with different PB fraction.} %\leavevmode
%\begin{tabular}[!h]{llllll}
%\hline
\tablehead{\colhead{$f_{\rm PB}$} & \colhead{RLO} & \colhead{RLO-PBs}     & \colhead{RLO-EBs}      & \colhead{RLO-MSBHs}  & \colhead{RLO-LBHs}}
\startdata %\hline
0.2  & $56\pm{}7$  & $39\pm{}6$           & $17\pm{}4$           & $6\pm{}2$   & $50\pm{}7$ \\
0.1  & $41\pm{}5$  & $22.5\pm{}3$      & $18.5\pm{}3$         & $9\pm{}2$          & $32\pm{}4$ 
%1.0  & 73  & 35          & 38           & 3          & 70 \\
%All$^{\rm b}$  & 244 & 137         & 107          & 42         & 202 \\
%\noalign{\vspace{0.1cm}}
%\hline
%\end{tabular}
%\begin{flushleft}
%\footnotesize{
 \enddata
\tablecomments{{\it Notes.} $f_{\rm PB}$ (column 1): PB fraction in the simulates SCs; RLO (column 2): number of all simulated RLO systems per SC;  RLO-PBs (column 3): number of PBs that undergo RLO;  RLO-EBs (column 4): number of EBs that undergo RLO; RLO-MSBHs (column 5): number of  RLO systems powered by MSBHs; RLO-LBHs (column 6): number of RLO systems powered by BHs with mass$<25$ M$_\odot$ for different metallicities and in total. \\
Both the SCs with $f_{\rm PB}=0.1$  and the SCs with $f_{\rm PB}=0.2$ have metallicity $Z=0.1$ Z$_\odot$. The tabulated numbers for the  SCs with $f_{\rm PB}=0.1$ come from Table~\ref{tab:tab4} but have been divided by two to match the number of runs with $f_{\rm PB}=0.2$  (we ran 200 simulations with $Z=0.1$ Z$_\odot$ and $f_{\rm PB}=0.1$ and 100 simulations with $Z=0.1$ Z$_\odot$ and $f_{\rm PB}=0.2$). All SCs have been simulated for 100 Myr.\\
The tabulated uncertainties are Poissonian errors.\label{tab:tab9}}
%\end{flushleft}
%\end{center}
\end{deluxetable}
%%%%%%%%%%%%%%%%%%%%%%%%%%%%%%%%%%%%%%%%%%%%%%%%%%%%%%%%%%%%%%%%%%%%%%%%%%%%%
%%%%%%%%%%%%%%%%%%%%%%%%%%%%%%%%%%% FIGURE 9 %%%%%%%%%%%%%%%%%%%%%%%%%%%%%%%%%%
\begin{figure}
\center{{
\epsfig{figure=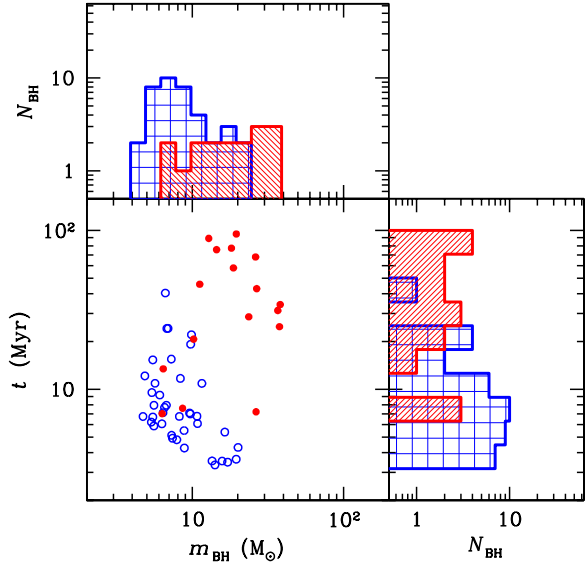,width=8.5cm} %was mBHvst_check.eps
}}
\caption{\label{fig:figA1}
Main window: mass of the BHs that power simulated RLO systems versus the time when the RLO phase starts for the 100 runs with $f_{\rm PB}=0.2$ and $Z=0.1$ Z$_\odot$. Red filled circles: RLO-EBs; blue open circles: RLO-PBs. The marginal histograms show the distribution of the masses of BHs that power RLO systems and the distribution of the times when the RLO phase starts ($t=0$ is the beginning of the simulation). In both marginal histograms, the red hatched histogram (blue cross-hatched histogram) refers to RLO-EBs  (RLO-PBs). }
\end{figure}
%%%%%%%%%%%%%%%%%%%%%%%%%%%%%%%%%%%%%%%%%%%%%%%%%%%%%%%%%%%%%%%%%%%%%%%%%%%%%%
 Table~\ref{tab:tab9} compares the main statistical properties of RLO systems in SCs with $f_{\rm PB}=0.2$ and with $f_{\rm PB}=0.1$, respectively. The number of RLO systems in SCs with  $f_{\rm PB}=0.2$ is considerably higher (by a factor of $\sim{}1.4$) than that of RLO systems in SCs with $f_{\rm PB}=0.1$. This is due to the much larger number of RLO-PBs in the SCs with higher $f_{\rm PB}$: the number of RLO-PBs in SCs with $f_{\rm PB}=0.2$ is a factor of $\sim{}1.7$ higher than that of RLO-PBs in SCs with $f_{\rm PB}=0.1$. 

In contrast, the number of RLO-EBs in the runs with $f_{\rm PB}=0.2$  remains substantially unchanged (see column 4 of Table~\ref{tab:tab9}) with respect to the runs with $f_{\rm PB}=0.1$. The number of RLO-MSBHs in the runs with $f_{\rm PB}=0.2$ is fairly consistent with that of RLO-MSBHs in the runs with $f_{\rm PB}=0.1$, within Poissonian uncertainties.

%The number of RLO-MSBHs in the runs with $f_{\rm PB}=0.2$ is slightly lower than that of RLO-MSBHs in the runs with $f_{\rm PB}=0.1$, but these numbers are fairly consistent within Poissonian uncertainties.

Fig.~\ref{fig:figA1} shows the distribution of BH masses and that of the times when the RLO phase starts for RLO-PBs and RLO-EBs in the SCs with $f_{\rm PB}=0.2$. RLO-EBs tend to host more massive BHs and to start the RLO phase later than RLO-PBs. This is fairly consistent with the results that we obtained for the SCs with $f_{\rm PB}=0.1$ (Figures~\ref{fig:fig1} and \ref{fig:fig2}).

In conclusion, the main effect of including a higher PB fraction in the simulated SCs is that the absolute number of RLO-PBs increases significantly. In contrast, the number of RLO-EBs is not affected by the initial binary fraction. We expect that the number of RLO-EBs depends more on the structural properties of the SCs (e.g. the central density and the timescale for core collapse). The other main results of this paper (i.e. the fact that  RLO-EBs statistically host more massive BHs/less massive donor stars and start the RLO phase later than RLO-PBs) are not affected by the PB fraction.
\end{appendix}

\end{document}